\RequirePackage{fix-cm}
\documentclass[smallextended]{svjour3}       
\smartqed  
\usepackage{amsmath}
\usepackage{graphicx}
\usepackage{xcolor}
\usepackage{comment}
\begin{document}

\title{Quantitative Modelling of Diffusion-driven Pattern Formation in microRNA-regulated Gene Expression
} 

\subtitle{}

\titlerunning{Pattern Formation by miRNA-regulated Gene Expression}        

\author{Priya Chakraborty, Sayantari Ghosh 
}


\institute{Priya Chakraborty$^*$ (corresponding author)\at
              Department of Physics, National Institute of Technology Durgapur \\
              \email{pc.20ph1104@phd.nitdgp.ac.in}          
              \and
           Sayantari Ghosh \at
              Department of Physics, National Institute of Technology Durgapur \\
              \email{sayantari.ghosh@phy.nitdgp.ac.in}
}

\date{Received: date / Accepted: date}

\maketitle

\providecommand{\keywords}[1]
{
  \small	
  \textbf{\textit{Keywords---  Epidemic models, Social media addiction, Complex networks, Bifurcation analysis, Pattern formation}}
}

\begin{abstract}
MicroRNAs are extensively known for post-transcriptional gene regulation and pattern formation in the embryonic developmental stage. We explore the origin of these spatio-temporal patterns mathematically, considering three different motifs here. For three scenarios, (1) simple microRNA-based mRNA regulation with a graded response in output, (2) microRNA-based mRNA regulation resulting in bistability in the dynamics, and (3) a coordinated response of microRNA (miRNA), simultaneously regulating the mRNAs of two different pools, detailed dynamical analysis, as well as the reaction-diffusion scenario have been considered and analyzed in the steady state and for the transient dynamics further.  We have observed persistent-temporal patterns, as a result of the dynamics of the motifs, that explain spatial gradients and relevant patterns formed by related proteins in development and phenotypic heterogenetic aspects in biological systems. Competitive effects of miRNA regulation have also been found to be capable to cause spatio-temporal patterns, persistent enough to direct developmental decisions. Under coordinated regulation, miRNAs are found to generate spatio-temporal patterning even from complete homogeneity in concentration of target protein, which may have impactful insights in choice of cell-fates.
\end{abstract} \hspace{10pt}

\section{Introduction}
Pattern formation in living systems dictates several long-term decision-making, in contexts starting from cell differentiation and morphogenesis, to development and phenotypic heterogeneity in bacteria. It started with the seminal work of A. Turing, opening up a new field of how the positional information is laid down in living cells, accounting for the mechanism of biological pattern formation with the help of the reaction-diffusion (RD) model \cite{turing1990chemical}. Examples of pattern formation in biological systems are countless. Patterning during embryonic development \cite{dinardo1994making}, morphogenesis \cite{lopes2008spatial}, organization of neural networks \cite{ermentrout1998neural}, and patterns on body i.e in wings of butterflies \cite{futahashi2012comprehensive}, pigments in fish body \cite{kondo2009animals} and many more. 
The existence of a chemical gradient in an embryo plays a key role in pattern formation in \textit{Drosophila melanogaster} \cite{borin1993egg,driever1988gradient}. In a hierarchy, the transition from simple gradient to complex patterning can be seen in giraffe \cite{koch1994biological}. The growth and development of tissues also affect the pattern formation in embryos \cite{kondo1995reaction}. Another prominent area that has been extensively explored by scientists in the last three decades is patterns produced by the reaction–diffusion models of Prey–predator \cite{banerjee2022stationary,menezes2022pattern,kumari2022controlling,jana2020self,liu2022study} interactions and other types of interacting ecological systems, Vegetation pattern formation \cite{sun2022dynamic,li2022bifurcation}, Activator-inhibitor system \cite{gierer1972theory,othmer1971instability}, feedback quenched oscillator system \cite{hsia2012feedback} and many more \cite{miyazako2013turing,mbopda2021pattern}. \\In spite of these explorations, the field of pattern formation by diffusible molecules in gene regulation is still underdeveloped. Recently, the field of transient pattern formation in gene regulatory dynamics has drawn the attention of the science community and some initial investigations have been performed \cite{chakraborty2023spatio,roy2022spatiotemporal,barbier2020controlling}. However, the diverse regime of post-transcriptional gene regulation remains completely unexplored, in this context. MicroRNAs (miRNA) are a class of small, non-coding RNA molecules majorly involved in post-transcriptional gene regulation \cite{bartel2004micrornas}. The presence of this single-stranded chain of nucleotides is found in plants, animals, and some viruses. In the case of humans and other mammals, miRNA targets nearly $60\%$ of the total mRNAs \cite{friedman2009most}. In plant morphogenesis, evidence of direct miRNA regulation is seen \cite{palatnik2003control}. In the post-transcriptional stage, miRNA binds with a messenger RNA (mRNA) and stops further translation, thus, protein synthesis. This sets a threshold in gene expression and dictates the level of stress or environmental fluctuation a cell can withstand.  In plant development, adaxial-abaxial polarity specification, meristem initiation, and auxin response factor genes are regulated by some particular types of miRNAs \cite{dong2022micrornas}. In mammals,  the importance of miRNA-regulation has been observed in development, apoptosis, adipocyte differentiation, neural cell fate, and hematopoiesis etc., \cite{bhaskaran2014micrornas,bissels2012micrornas}. Dysregulation of miRNA is found in disease formation like cancer \cite{calin2006microrna} and in neuronal disorders \cite{kosik2006neuronal}, tumor progression/regression, cholesterol, glucose homeostasis, etc. As drivers of post-transcriptional regulation, a single miRNA species is capable of regulating several different mRNAs; during post-embryonic development, this \textit{coordinated response} in miRNA regulation for multiple pools of mRNAs plays a leading role in achieving proper developmental timing and  cell differentiation \cite{chen2009small,jones2006micrornas}.
\\Though miRNAs, as an important post-transcriptional regulator in gene expression dynamics, has an evident role in causing spatial heterogeneity (like development), reaction-diffusion-based pattern formation for miRNA dynamics has been a little explored area. Some of the recent works are related to pattern formation and gene amplification during Drosophila oogenesis\cite{ge2015regulation}, experimental evidence of transient focal ischemia by middle cerebral artery occlusion of rats by miRNA expression\cite{jeyaseelan2008microrna}, embryo pattern formation at the beginning of zygote \cite{armenta2017arabidopsis} etc. Though, there is a large scope of the study, specifically exploring quantitative models of pattern formation by miRNA-regulated genetic motifs and respective diffusible protein molecules with/without environmental fluctuations. Different transient or steady state patterns may emerge in different environmental conditions and elaborative studies on them will help in understanding phenomena like cell-fate decision-making, biochemical signaling, and many more. 
\\Here, in this paper we have explored three different motifs of miRNA-mediated mRNA regulation, their dynamical behaviors, and majorly the pattern formation by the motifs in a diffusible cellular environment for different initial conditions. Instead of studying the dynamics of a single cell, we consider a collective cell arrangement to explore the genetic motifs which are closer to experimental scenarios. The paper has been organized in the following way: in Section $2$, we have explored the miRNA-mediated threshold gene expression in a diffusible cellular environment; in Section $3$ we have explored a binary gene expression by post-transcriptional regulation of miRNA in terms of steady-state dynamics and reaction-diffusion model. In Section $4$, a model of coordinated response by miRNA, both in steady state and reaction-diffusion model for a collective cellular array are explored and reported. Finally, in Section $5$, we conclude with some discussion and the future scope of explorations.      
\section{MicroRNA can create Spatial Thresholds in Protein Response}
Via post-transcriptional regulation, miRNA controls the gene expression in a number of important aspects like setting thresholds for gene expression, suppressing fluctuations, filtering out transient signals, and many more. Generally, miRNA binds to target mRNA with imperfect complementarity, producing an mRNA-miRNA bound complex which can either degrade or remain inactive for a long time. Thus, the translation is restricted resulting in null protein output. 
\subsection{Model Formulation}
\begin{figure}
    \centering
    \includegraphics[width=\textwidth]{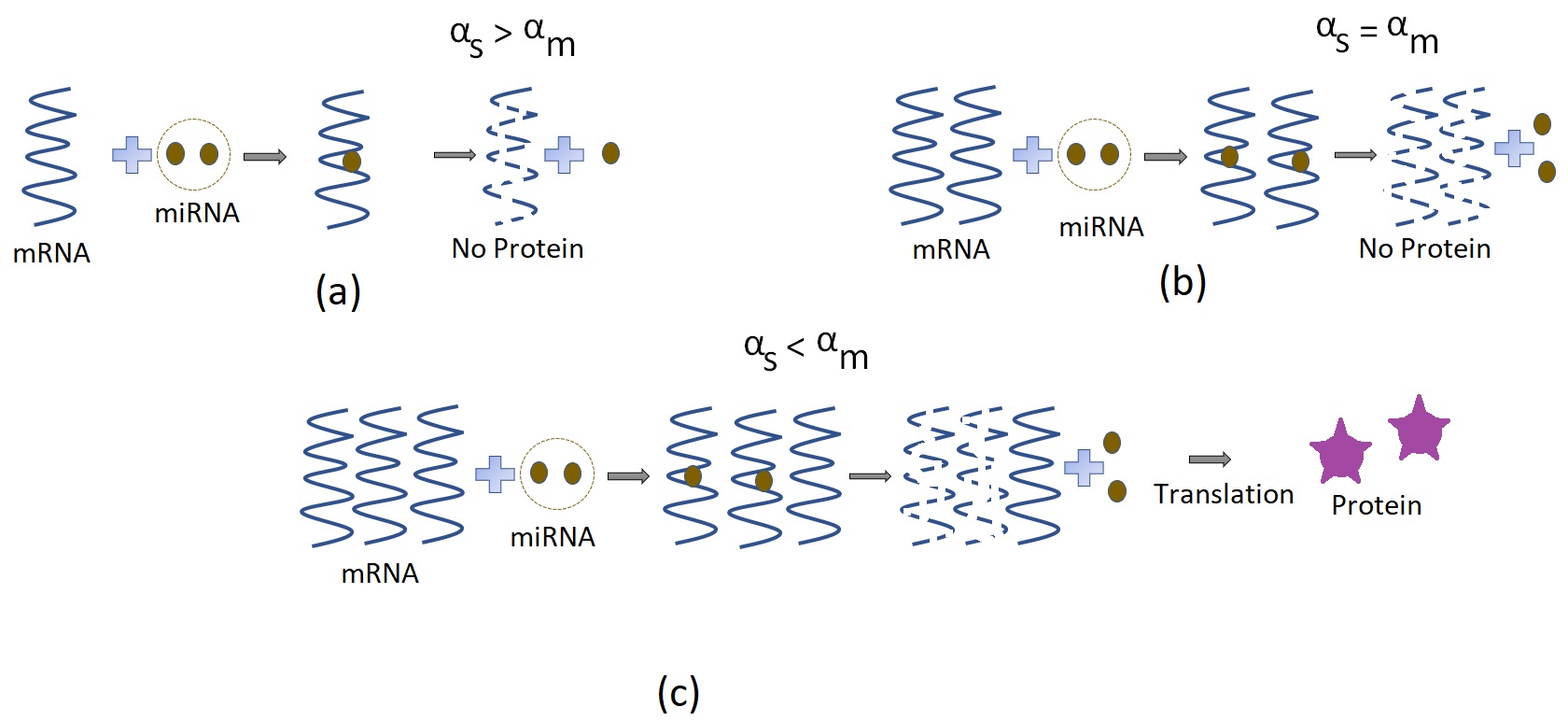}
    \caption{Schematic diagram for threshold linear response in gene expression by the post-transcriptional regulation of miRNA. (a) If the transcription rate of miRNA $\alpha_s$, is greater than the transcription rate of mRNA $\alpha_m$, i.e., $\alpha_s>\alpha_m$, no active mRNA will remain for further translation. Thus no protein will synthesize. (b) If $\alpha_s=\alpha_m$, all mRNA will be targeted by miRNA, and no active mRNA will remain to produce protein further. (c) If $\alpha_s<\alpha_m$, some mRNA will eventually remain active for further translation and protein synthesis.}
    \label{thresold}
\end{figure}
In this model (referred as \textit{Model 1} in the rest of the manuscript), let us consider a class of miRNA that have a single mRNA target at steady state. Let $\alpha_s$, $\alpha_m$ be the rates of transcription of miRNA and mRNA respectively. Let $\beta_s$ and $\beta_m$ be the rates of natural degradation of miRNA (say $S$) and mRNA (say $m$), while $k$ is rate of complex formation between miRNA and mRNA, and degradation of the complex further. The rate of protein (say $P$) production is considered as $\alpha_p$, linear with the active mRNA level, and the natural degradation is considered as $\delta$. Now, the set of differential equations representing the above is given by:  
\begin{eqnarray}\label{thrsld}
    \frac{dS}{dt}=\alpha_s-\beta_s\;S-k\;m\;S\nonumber\\
    \frac{dm}{dt}=\alpha_m-\beta_m\;m-k\;m\;S\\
    \frac{dP}{dt}=\alpha_p\;m-\delta\;P\;\nonumber
\end{eqnarray}
Now, in presence of spatial heterogeneity, we consider a two-dimensional sheet of cells and further diffusion of the protein, to mimic the tissue layer formation of cells. The position is discretized as $(x_i,y_i),i\in ({1,200})$, we have considered a $200\times200$ arena of two-dimensional cellular arrangement. This consideration is valid for all the models, explored in this paper. The synthesized protein is allowed to diffuse through the cell sheet in a no-flux boundary condition.  Let $D_P^x$ and $D_P^y$ be the diffusion coefficients along the two directions $x$ and $y$ here. Then the Eqn. \ref{thrsld} will change to  
\begin{eqnarray}
    \frac{dS}{dt}=\alpha_s-\beta_s\;S-k\;m\;S\nonumber\\
    \frac{dm}{dt}=\alpha_m-\beta_m\;m-k\;m\;S\\
     \frac{\partial P(x_i,y_i,t)}{\partial t}=\alpha_p\;m-\delta\;P+D_P^x\;\frac{\partial^2P}{\partial x^2}+D_P^y\;\frac{\partial^2P}{\partial y^2}\nonumber
\end{eqnarray}
Further considering isotropic diffusion (equal diffusion in both directions), we have $D_P^x=D_P^y=D_P$ (say).
\subsection{Results}
\begin{figure}
    \centering
    \includegraphics[width=\textwidth]{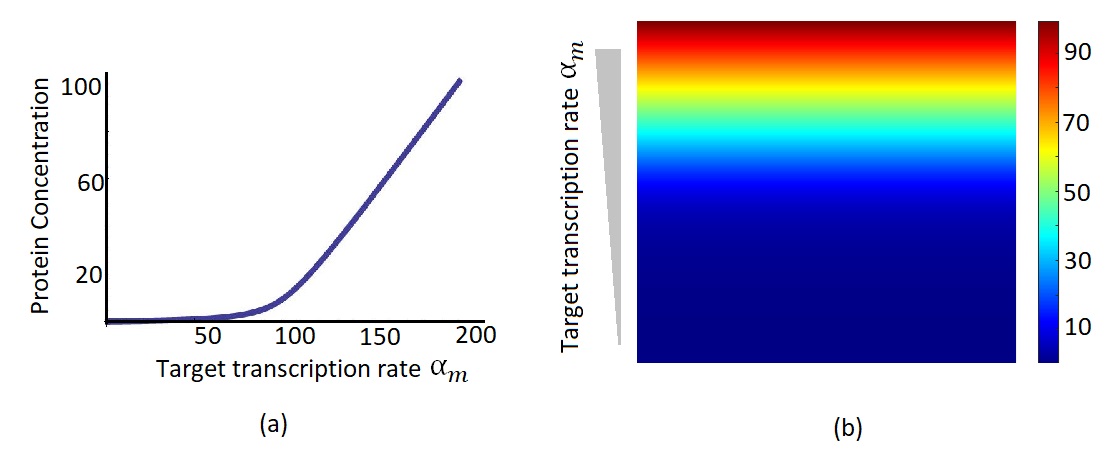}
    \caption{Threshold linear response in miRNA-based gene expression. (a) The concentration of protein $P$ remains $0$ initially, crossing after a threshold value of $\alpha_m$ (nearly at $\alpha_s=\alpha_m$) the protein starts synthesizing and increases linearly with an increase in $\alpha_m$. (b) For two dimensional array of cells, threshold linear response in reaction-diffusion environment. The target transcription rate of mRNA is varied in $y$ direction and a threshold response is shown in the output. Parameter values for both the figures are $\alpha_s=100,\;\beta_s=1,\;\beta_m=1,\;k=1,\;\alpha_p=1,\;\delta=1.$ $\alpha_m$ is varied from $1$ to $200$, in $x$ axis for (a), and in $y$ axis in for (b). }
    \label{Threshold-linear}
\end{figure}

To mathematically establish the linear threshold response, we plot the synthesized protein concentration in $y$ axis, for a fixed value of miRNA transcription rate $\alpha_s$ at $100$, and for an increase in mRNA transcription rate $\alpha_m$ in $x$ axis in the range of $0$ to $200$. Protein is considered to be synthesized linearly with mRNA, as shown in Eqn. \ref{thresold}. Hence, the level of protein concentration is actually equivalent to the level of mRNA available for translation.\\
Now if the rate of mRNA transcription ($\alpha_m$) is lower than that of the rate of miRNA synthesis ($\alpha_s$), as soon as the mRNA is produced after transcription, the miRNA binds to it making a complex that stops further translation of mRNA, thus, as a result, no protein synthesizes. Conversely, if $\alpha_s$ is less than the rate of mRNA synthesis $\alpha_m$, then some mRNA cannot be targeted due to a lack of miRNA, which can be further translated to proteins. This sets a threshold response in gene expression at $\alpha_s \approx \alpha_m$. A schematic diagram of these conditions is shown in Fig. \ref{thresold} (a)-(c). The position of the threshold depends upon the value of $\alpha_s$ with no change in slope and the target expression level is comparable to the difference between $\alpha_s$ and $\alpha_m$. The threshold linear response behavior is shown in Fig. \ref{Threshold-linear}(a). It is clear that the kink is near the region of $\alpha_s \approx \alpha_m$ and a linear increase in protein level is seen afterward. A spatially-extended response showing threshold-linear behavior can be seen in a two-dimensional cellular array, of 40,000 cells in Fig. \ref{Threshold-linear}(b).\\
This controlled initiation and maintenance of a spatial gradient, as a result of the considered post-transcriptional regulation, in the production of the output protein is quite common in different phases of development. Further, the capacity of the stress response and/or fluctuation sensitivity of a cell cluster can be captured by this miRNA-based mRNA regulation. During cell fate decisions in development, when a protein is only expressed above a certain threshold, this miRNA-based regulation can be extremely effective in setting the threshold level in cell systems \cite{shu2019opposing}. A spatially ultrasensitive all-or-none gradient dictated by the miRNA threshold can be generated, resulting in the spatial expression pattern of the targets. In terms of wet-lab experiments, thankfully, the transcription rates of miRNA and mRNA can be easily dynamically controlled by regulating the activity at their promoters, rather than changing degradation rates, either single or coupled, which are biologically difficult to tune, makes it synthetically feasible to. 

\section{Bistable gene expression \& Pattern Formation by miRNA regulation}
\begin{figure}
    \centering
    \includegraphics[width=\textwidth]{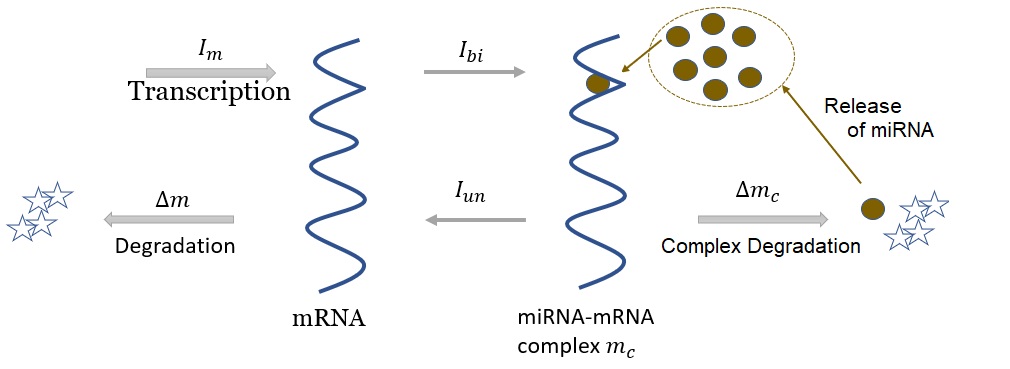}
    \caption{Schematic diagram describing miRNA-based mRNA regulation. After transcription at the rate of $I_m$, the target mRNA can do further translation for the final protein product or mRNA can degrade at a rate of $\Delta m$. Meanwhile, miRNA binds with free mRNA at a rate of $I_{bi}$ to form miRNA-mRNA complex $m_c$. The complex can unbind at a rate $I_{un}$ or can degrade at a rate of $\Delta m_c$. The protein expressed gives a positive feedback to the mRNA transcription (not shown here).}
    \label{schematic2}
\end{figure}
In this model (referred as \textit{Model 2} in rest of the manuscript), we consider bistability caused by miRNA-regulated gene expression. Inspired by a recent model \cite{bose2012origins}, we study a miRNA regulated protein synthesis, along with a non-cooperative positive feedback loop in the transcriptional regulation. First, analysing the model from deterministic perspective, we extend the study for a cellular array of $2$ dimensional sheet, allowing the protein diffusion for a closer-to-reality investigation, considering the tissue formation in biological systems.

\subsection{Model Formulation:}
Let us consider, a single miRNA species in regulating the translation of the mRNA for protein $U$. The mRNA is either free $(m)$ or has formed miRNA-mRNA complex $m_c$.  The total concentration of the miRNA is considered $mi_{Total}$, and the free miRNA is denoted as $mi$ such that
\begin{equation}
  mi_{Total}=mi+m_c  
\end{equation}
The half-life of miRNA (which is experimentally determined as $28$ to $220$ hour)  is roughly $2$ to $20$ fold longer than the half-life of mRNA (generally 10 hours) \cite{zhang2012microrna}. So, the total concentration of miRNA can be considered constant in gene expression dynamical equations. The natural degradation constant for mRNA and miRNA is taken as $\Delta m$ and $\Delta mi$. The transcription rate by which the target mRNA is being produced is taken as $I_{m}$, and the rate constants for binding and unbinding of mRNA and miRNA to produce the complex is taken as $I_{bi}$ and $I_{un}$. Demonstrating two possibilities, the bound complex can either cause degradation of the mRNA at rate $\Delta m_c$, releasing the miRNA, or can break up into free mRNA and free miRNA. A non-cooperative auto-activation of the target gene is considered with a rate constant $\gamma_m$ with an equilibrium dissociation constant $k$. The rate of total protein production is considered as $I_u$, and the degradation of protein $U$ is considered as $\Delta u$. A schematic diagram of this model consideration is shown in Fig. \ref{schematic2}. The differential equations describing the model are given by:
\begin{eqnarray}
 \frac{dm}{dt}= I_m+\frac{\gamma_m\;U}{k+U} -I_{bi}\;m\;mi+I_{un}\;m_c-\Delta m\;m\nonumber\\
 \frac{dm_c}{dt}= I_{bi}\;m\;mi-I_{un}\;m_c-\Delta m_c\;m_c\\
 \frac{dU}{dt}= I_u\;m-\Delta u\;U\nonumber
\end{eqnarray}
In the steady state, all the rates of change are equal to zero, and modified equations are given by

\begin{eqnarray}\label{abbriviatn}
    I_m+\frac{\gamma_m\;U}{k+U}-I_{bi}\;m\;mi+I_{un}\;m_c-\Delta m\;m=0\nonumber\\
    \beta^*=\Delta m_c\;mi_{Total},\;\;\;\;\;
    m_c=\frac{m\;mi_{Total}}{m+\lambda_a},\;\;\;\;\;\; \lambda_a=\frac{I_{un}+ \Delta m_c }{I_{bi}}\\
    U=\frac{I_u}{\Delta u}\;m\nonumber
\end{eqnarray}
Further calculations demonstrate that the steady-state solution for protein $U$ will be given by:
\begin{equation}\label{binary}
\delta+\frac{U\;\alpha}{k+U}-\frac{\phi\;U}{\lambda\;+U}-U= f(U)=0  ,  
\end{equation}
where $\delta=\frac{I_m}{\Delta_\alpha}$, $\alpha=\frac{\gamma_m}{\Delta_\alpha}$, $\phi=\frac{\beta^*}{\Delta_\alpha}$, $\lambda=\lambda_\alpha\;\frac{I_u}{\Delta u}$, and $\Delta_\alpha=\frac{\Delta m\;\Delta u}{I_u}$.
\subsection{Stability analysis and Bifurcation}
To determine the equilibrium points that were biologically feasible, we equate the system Eqn. \ref{binary} to zero. Graphically, Fig. \ref{lsa}(a) shows that $\textit{f(U)}$ may intersect $U$ axis in \textit{three} points for certain parameter values, indicating maximum \textit{three} relevant solutions of the system; however, the function can also change its slopes to result into a single solution for some other parameter regime. This indicates the presence of bifurcation in the dynamics.\\
Theoretical proof of the possible bifurcations can be approached through the discriminant of the cubic polynomial resulting from the system Eqn. \ref{binary}. We proceed to study the discriminant of the cubic system:
\begin{equation}\label{discrimnt}
     U^3+X\;U^2+Y\;U+Z=0,
\end{equation}
where the coefficients of the polynomial are represented in terms of system parameters as:
\begin{eqnarray}
    X=(\lambda\;+\;k\;+\;\phi\;-\;\delta\;-\alpha), \nonumber\\
    Y=( \lambda\;k\;+\;\phi\;k-\alpha\;\lambda-\delta\;k-\delta\;\lambda),\nonumber \\
    Z=-\;\delta\;k\;\lambda \nonumber
\end{eqnarray}
When plotted graphically, in Fig. \ref{lsa}(b), for a range of $\alpha$ value, the discriminant $\Delta$ is positive, and the system has \textit{three} real positive solutions.\\
For the system to be bistable, for all three roots of Eqn. \ref{discrimnt}, we must have a positively invarient set, $ \Omega=\{U \in \mathcal{R}_+ : U \geq 0\}$, which implies:
\begin{equation}\label{discriminant}
    \Delta\;=\;18XYZ\;-\;4X^3Z\;+\;X^2Y^2-4\;Y^3-27\;Z^2\geq 0
\end{equation}
Boundaries of this condition, $\Delta=0$, indicates existence of bifurcation points where two solutions collide and annihilate each other, giving rise to \textit{saddle-node bifurcations}. \\
As expected in this kind of bifurcation, out of the three possible solutions, further analysis of linear stability determines the existence of $two$ stable solution and one unstable solution. The protein $U$ shows bistability for a range of $\alpha$ values ($\alpha_{low}$ to $\alpha_{high}$) and these two points are called \textit{lower and upper bifurcation points }($\alpha_{low}$ and $\alpha_{high}$, respectively marked in Fig. \ref{lsa}(c)).  Here, while increasing the parameter $\alpha$, we find no fixed point exists as a continuation of low synthesis steady state beyond  $\alpha=\alpha_{high}$, as two fixed points (unstable points shown using the dotted line and low synthesis stable point, shown using the solid line) approach each other and annihilate at $\alpha_{high}$. A similar statement can be made for $\alpha_{low}$, where the high synthesis state collides and annihilates with the unstable point. Temporal behavior of the system, starting from different initial conditions, shown in Appendix, Fig. \ref{flowline} also supports the bistable nature of the system for the given range of parameter values that satisfies $\Delta \geq 0$.  Now, let us elaborate on the bistable nature of the system wrt. different parameters.
\subsubsection{Bistability of protein $U$: Parameter Variation} 
\begin{figure}
    \centering
    \includegraphics[width=\textwidth]{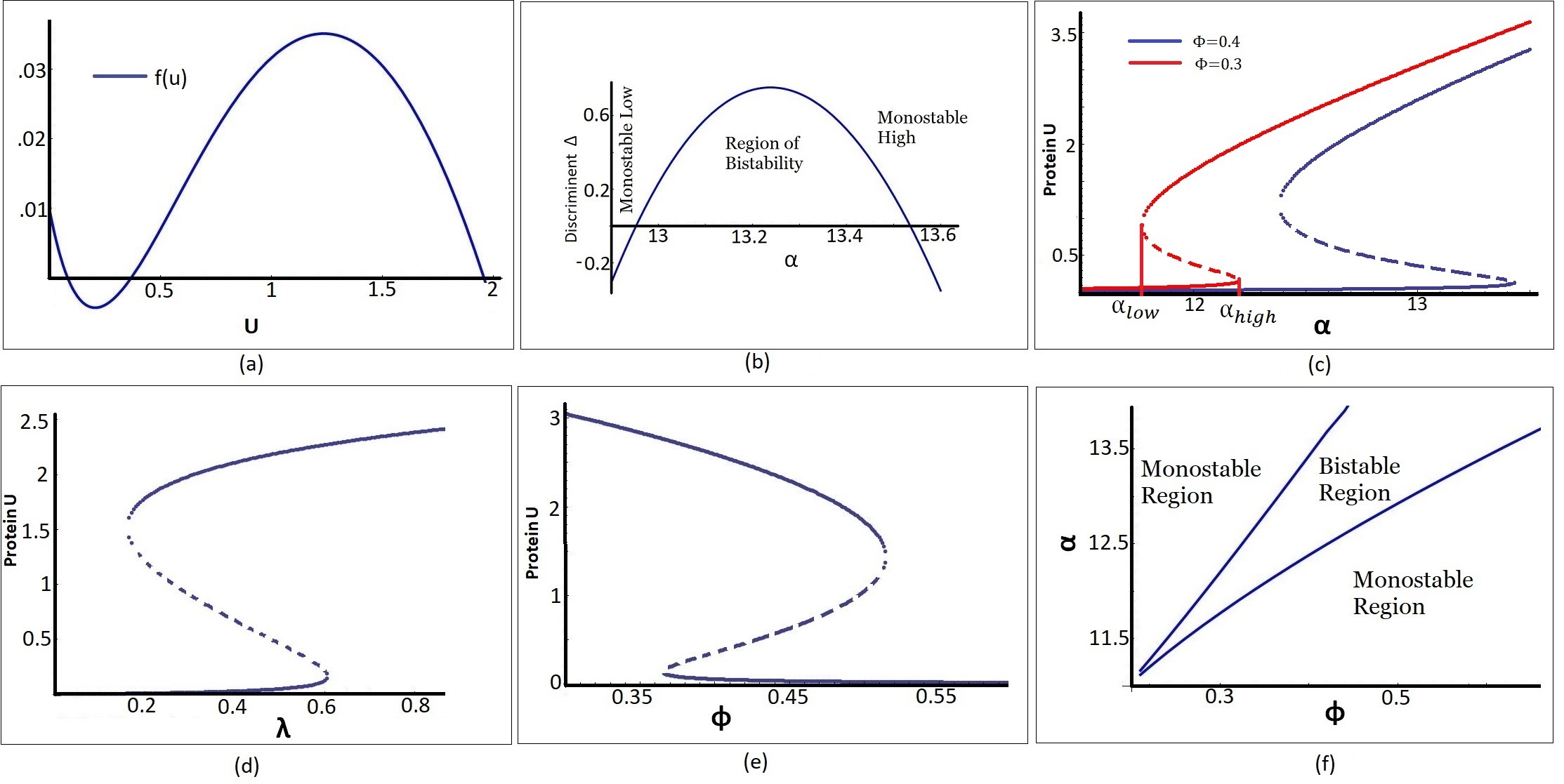}
    \caption{Linear stability analysis of the motif. (a) $f(U)$ curve intersect $U$ axis in 3 different point. Parameter values are $\alpha=13.3,\;\delta=0.01,\;k=10,\;\lambda=0.6,\;\phi=0.3$. (b) Plot of discriminant $\Delta$ with respect to parameter $\alpha$. Bistability is achieved in the system when $\Delta \geq 0$. Monostable high, low, and region of bistability are respectively marked in the figure. Parameter values are $\lambda=0.6,\;\phi=0.3,\;k=10,\;\delta=0.01$. (c) Bistability diagram of protein $U$ wrt. parameter $\alpha$, blue curve is for $\phi=0.4$ and red curve is for $\phi=0.3$. (d) Bistability diagram of protein $U$ wrt. parameter $\lambda$. Parameter values are $\delta=0.01,\;k=10,\;\phi=0.3,\;\alpha=13.5.$ (e) Bistability curve of protein $U$ wrt. parameter $\phi$ for $\alpha=13$. (f) Phase space diagram of protein $U$ in $\alpha$-$\phi$ space. Rest of the parameter values for (c), (e), (f) are $\delta=0.01,\;k=9,\;\lambda=0.6$.}
    \label{lsa}
\end{figure}

In Fig. \ref{lsa}(c), the system is explored in terms of bifurcation,  tuning parameter $\alpha$, for two different values of $\phi$. For $\phi=0.4$, we can see that protein $U$ shows bistability for a range of $\alpha$ value, and the range of the bistable region shifted to a high value of $\alpha$ (blue curve) than that for $\phi=0.3$ (red curve), along with an increase in the region of bistability. The phase space plot wrt. $\phi-\alpha$ in Fig. \ref{lsa}(f),  the respectively marked monostable and bistable regions, also indicate a similar behavior. \\Biologically, this switch-like response, as a result of the emergence of a saddle-node bifurcation in the system here, is very effective in introducing a memory in the system between bifurcation points $\alpha_{low}=11.8$ and $\alpha_{high}=12.2$  (for $\phi=0.3$, red curve, Fig. \ref{lsa}(c)). While increasing the tuning parameter (say $\alpha$), the output protein concentration remains low and steady upto $\alpha=\alpha_{high}$, and a sudden jump to higher concentration value is seen after the point $\alpha_{high}$. But, in time of decreasing the tuning parameter $\alpha$, the protein concentration will not come to its low value for $\alpha=\alpha_{high}$ but chooses to retain its high state upto $\alpha=\alpha_{low}$ and a sudden jump to low concentration is seen after this point. A memory effect has been reported, for the range of $\alpha$ value from $\alpha_{low}$ to $\alpha_{high}$, the system tries to retain its previous high/low expression state, despite the fluctuation in its tuning parameter value. This range of parameter values, for which the system shows the memory effect, accounts for the robustness of the switch response in the biological systems. Physically two different protein concentrations can coexist here, depending upon the forward/backward operation mode of the system. Similar responses observed for two more parameters $\lambda$ and $\phi$ are also  reported in Fig. \ref{lsa}(d),(e).  \\
In the case of development, cell fate decision-making, this switch response plays a major role in biological systems. In miRNA-regulated post-transcriptional gene regulation resulting in a bistable behavior is found to regulate the cellular decision-making broadly in precise tissue boundary formation\cite{li2021microrna}, cell fate decision making \cite{tian2016reciprocal,tian2019modeling} and many more. The bistable behavior, of our model, thus should be studied in similar scenarios of biological systems with spatially extended model.

\subsection{Results: Reaction-Diffusion Model}
Here, we have explored the behavior of this motif in a two-dimensional cellular array to see the spatiotemporal response of the dynamics. In the presence of diffusion of $U$, in a two-dimensional cellular array of $200\times200$ cells, (similar as considered in Model 1) Eqn. \ref{binary} changes to
\begin{equation}
 \frac{\partial U(x_i,y_i,t)}{\partial t}=\delta+\frac{U\;\alpha}{k+U}-\frac{\phi\;U}{\lambda\;+U}-U +D_U^x\;\frac{\partial^2U}{\partial x^2}+D_U^y\;\frac{\partial^2U}{\partial y^2}  
\end{equation}
where, $D_U^x$ and $D_U^y$ is considered as the diffusion coefficient in the $x$ and $y$ direction. In our study, we have considered isotropic diffusion (thus $D_U^x\;=\; D_U^y\;=\; D_U$) of the protein in a no-flux boundary condition. We have further considered different initial conditions, remembering the variability in pattern formation in different kinds of systems. A scaling term $k_1$ has been added to the respective distribution functions, representing the two-dimensional protein distribution in the cellular sheet, randomized via the term $\xi$, which picks any number randomly between $0$ to $1$. On a common theme, with time evolution, a spatial pattern arises as a result of bistability and protein diffusion for all these initial conditions; however, the generated patterns have distinct natures.
\subsubsection*{Random initialization:}
\begin{figure}
    \centering
    \includegraphics[width=0.8\textwidth]{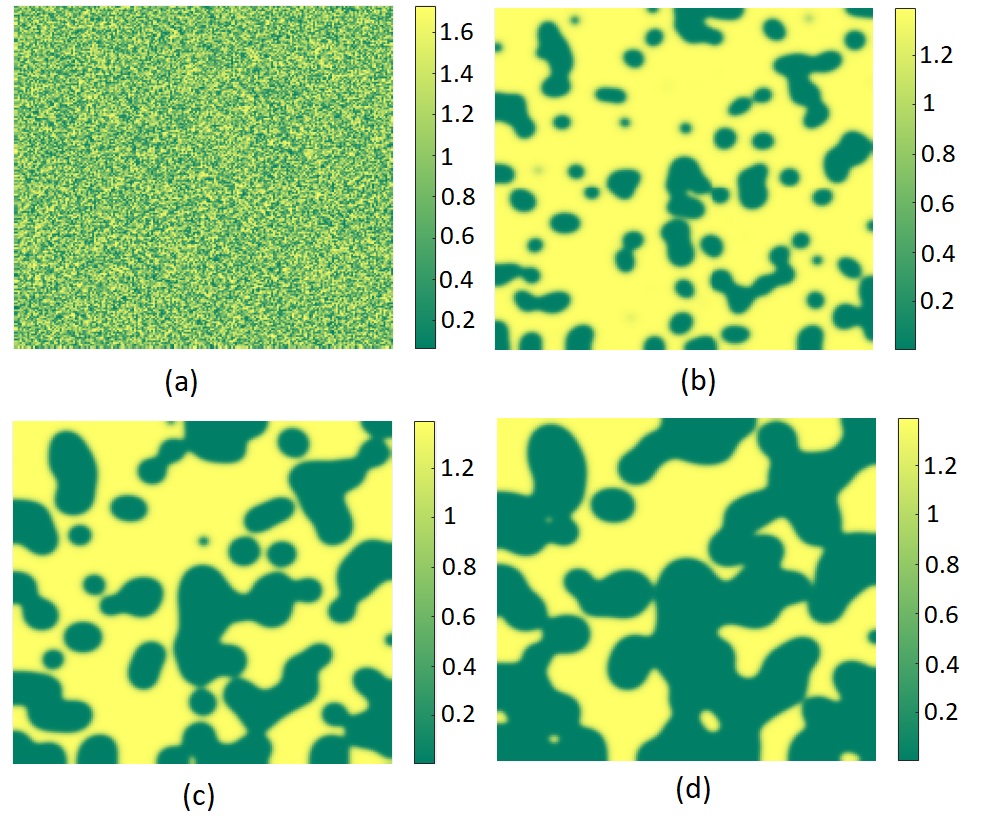}
    \caption{Diffusion in two dimensions for randomly distributed initial condition. Parameter values are $\alpha=13.1,\;\phi=0.3,\;\lambda=0.6,\;\delta=0.01,\;k=10,\;k_1=1.8$. Diffusion coefficient $D_U$ is taken as $0.05$. Snapshots are taken after time (a) 1 (b) 250 (c) 500 (d) 700.}
    \label{2drand}
\end{figure}

To begin with, let us consider,
\begin{equation*}
    U_{initial}(x_i,y_i,0) =k_1\;\xi(0,1)
\end{equation*}
As shown in Fig. \ref{2drand}(a)-(d), starting from a completely random initial condition, the system quickly achieves low or high expression states. Quickly the system converts to a mixture of some islands of high expression state and some islands of low expression state. Diffusion-driven instability causes the high-expression steady states to convert into the low-expression steady state, and thus the islands of low-expression states get bigger with time, getting connected to each other. This transient response causes a transient pattern formation in the system. As time progresses, the system will eventually converge to the state with greater stability (here, the low $U$ expression state).
\subsubsection*{Positive exponential initial condition:}
\begin{figure}
    \centering
\includegraphics[width=0.8\textwidth]{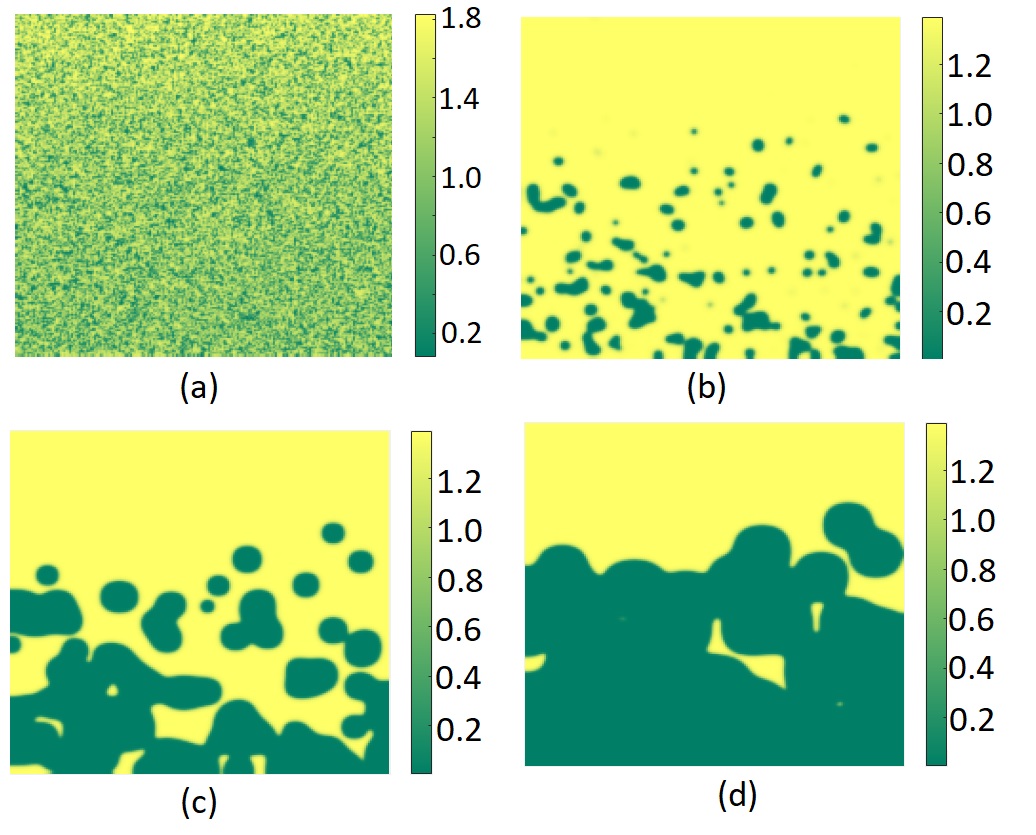}
    \caption{Diffusion in two dimensions with positive exponential initial condition. Snapshots are taken after time (a) 12 (b) 225 (c) 1000 (d) 2000. Parameter values are $\alpha=13.1,\; \phi=0.3,\;\lambda=0.6,\;\delta=0.01, k=10,\;k_1=2,\;k_2=0.00001.$ Diffusion coefficient $D_U=0.02$.}
    \label{expmodel2}
\end{figure}

A gradient of some inducer, either naturally occurring or experimentally generated, may cause these exponential initial conditions that give rise to distinctive threshold-based patch patterns. For this, we consider initial condition:
\begin{equation*}
U_{initial}(x_i,y_i,0) = k_1\;\xi(0,1)\;exp(k_2\;x_i^2) \end{equation*}
Exponential initial condition has also been tested and the corresponding spatio-temporal pattern formation is shown in Fig. \ref{expmodel2}(a)-(d). The visibly imperceptible gradient in initial values in Fig. \ref{expmodel2}(a) causes transient islands of low protein state (Fig. \ref{expmodel2}(c)), which grows further and creates a clear boundary of two expression states in later times (Fig. \ref{expmodel2}(d)), before converging to homogeneous state. At any intermediate state, any decision taken based on this cellular response, can propagate through the downstream pathways creating significant future effects.

\subsubsection*{Stochastic periodic initial condition: }
\begin{figure}
    \centering
    \includegraphics[width=\textwidth]{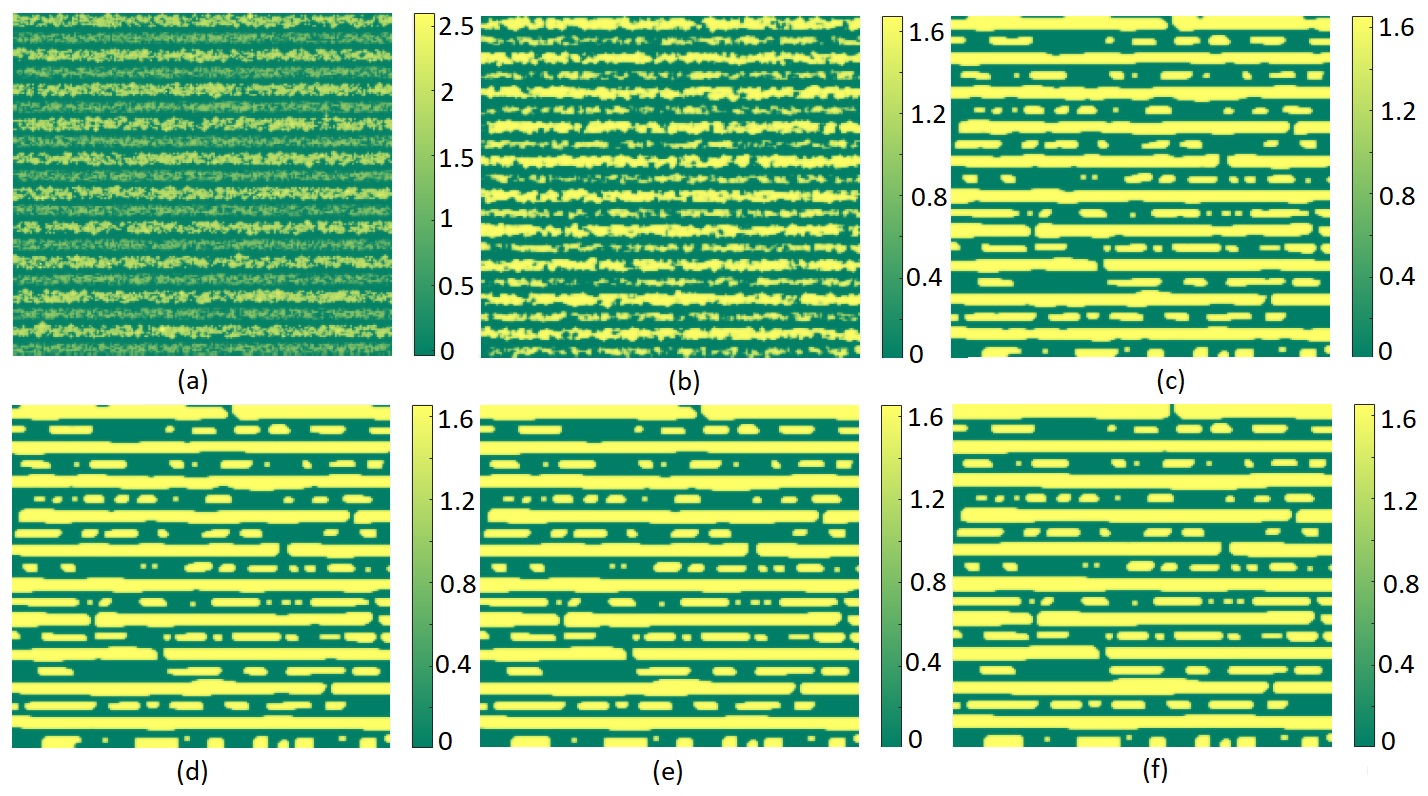}
    \caption{Diffusion in two dimensions with stochastic periodic initial condition. Snapshots are taken after time (a) 30 (b) 80 (c) 1350 (d) 2000 (e) 3000 (f) 5000. Diffusion constant is taken as $D_U=0.01$. Rest of the parameter values are $\alpha=13.2,\;\phi=0.3,\;\lambda=0.6,\;\delta=0.01,\;k=10,\;k_1=1.5,\;k_2=10. $}
    \label{twodsin}
\end{figure}

The periodic pattern formation is very common in biological systems (including zebra, cuckoo, zebrafish, etc. in animals and fern, Aloe Polyphylla, etc., in plants). Considering the importance of periodicity in biological systems, we explore the effects and behavior of miRNA-mediated mRNA regulation and, thus, protein synthesis in a bistable dynamical picture with following initial condition:
\begin{equation*}U_{initial}(x_i,y_i,0)=k_1\;\xi(0,1)sin(\frac{\pi\;x_i}{k_2})\end{equation*}
Starting from a stochastic sinusoidal initial condition, the time evolution and a pattern in output as shown in Fig. \ref{twodsin}. A transient evolution is shown in the figure; this might be important in the case of the developmental aspect as the output shows the pattern is very persistent. For this pattern formation, a video of the simulation can be seen here, which shows that the pattern becomes almost invariant with time\footnote[1]{Link for the video is: https://youtu.be/C4zovgxtCpQ}.  
\subsubsection*{Uneven stripes with tangential initial condition:}
\begin{figure}
    \centering
    \includegraphics[width=\textwidth]{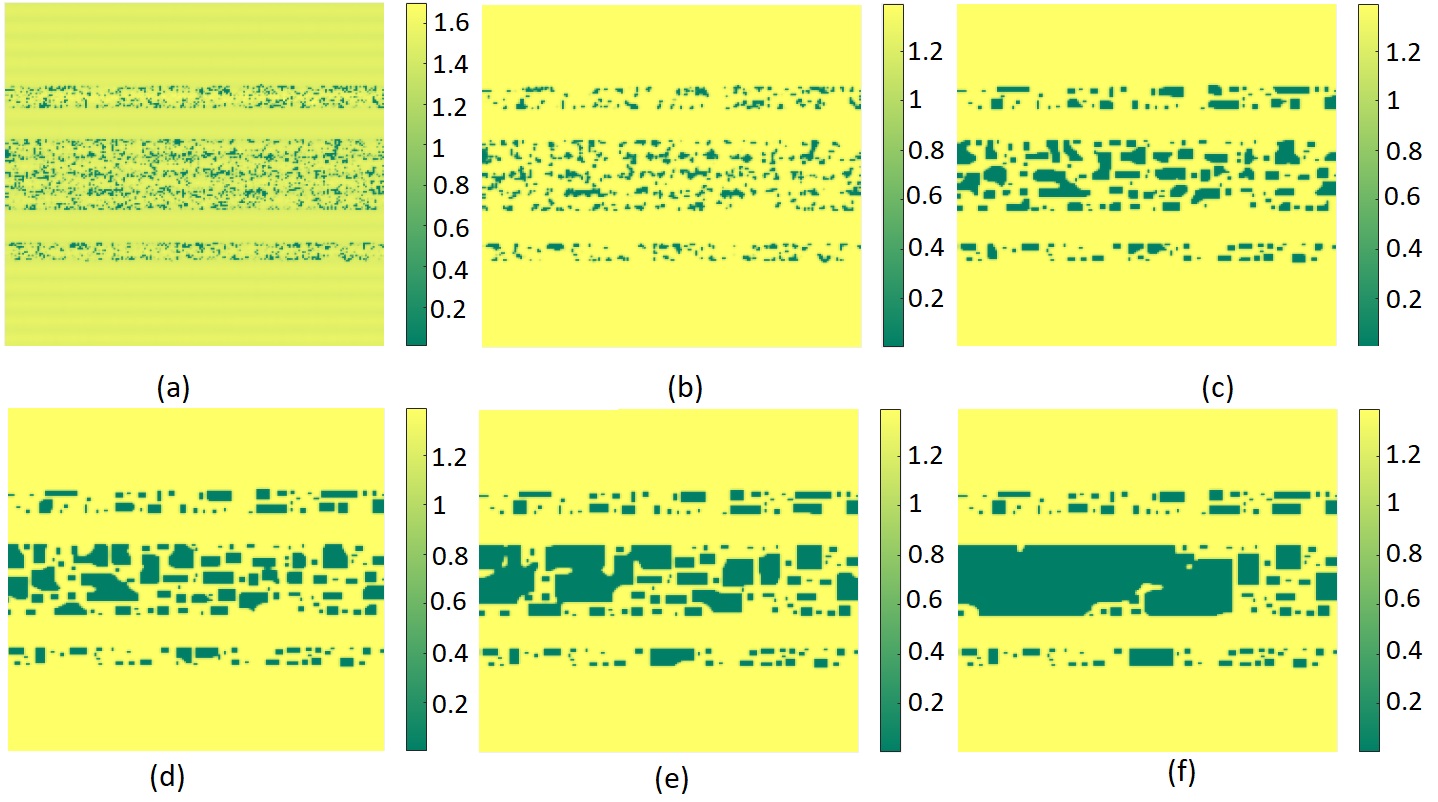}
    \caption{Diffusion in two dimensions with tangential initial condition. Snapshots are taken after time instants (a) 75 (b)240 (c) 1200 (d)2120 (e) 4400 (f) 9000. Different parameter values are $\alpha=13.1, \phi=0.3, \lambda=0.6,\delta=0.01,\; k=10,\;k_1=0.6,\;k_2=10,\; D_U=0.005$. }
    \label{2dtan}
\end{figure}

The final initial condition that we report is given by:
\begin{equation*}
    U_{initial} =k_1\;\xi(0,1)\;tan(\frac{\pi\;x_i}{k_2})
\end{equation*}
We choose this specific initial condition to look for heterogeneous stripe width. In the two-dimensional cellular arena as given above, the spatio-temporal evolution from this initial condition and respective pattern formation is shown in Fig. \ref{2dtan}(a)-(f). Spatial patterns, evolving with time, give rise to island sizes of large variability in a long time limit.
\subsection{Spatio-temporal patterns, initial conditions, and their steadiness:}
Considering the immense diversities and possibilities in biological systems, we have considered different initial conditions for pattern formation to explore here. It is important to note that though the patterns we are getting are transient spatio-temporal but are very persistent too. Some of our simulations eventually ended in the final low expression steady states (like Fig. \ref{2drand}, \ref{expmodel2}), but some other simulations are quite away from converging to a particular steady state (like Fig. \ref{twodsin}, \ref{2dtan}). It is possible in a cellular environment, in the presence of different intracellular and extracellular signals and activity, the patterns get steady. Even this long transient behavior is also very impactful for cellular decision-making. 
\section{Coordinated Response of miRNA: steady-state analysis and spatiotemporal pattern formation by sharing miRNA between two mRNA pools}
Now, let us proceed further for an extension of Model 2, considering that the pool of miRNA is now simultaneously accessible for two pools of different mRNAs. This is a commonly occurring phenomenon called the coordinated response of miRNAs, where the presence of two mRNA pools mutually helps each other for protein synthesis by sharing the available active miRNA. If the available miRNA pool is greater than each of the mRNA pools individually, and also greater than the sum of the two mRNA pools in number, then no mRNA from either pool will translate further. Here, the post-transcriptional regulation via miRNA is becoming prominent because of the dominant nature of the number game in miRNA-based regulation. 
The situation is similar and straightforward when the total miRNA is less than the individual, when both mRNAs can proceed for translation. However, the coordinated response is more fascinating when the miRNA pool is greater than each of the individual mRNA pools, but less than the sum of two mRNA pools. As the miRNA pool now exceeds each mRNA pool in number, no translation is expected due to miRNA-mRNA pairing, thus no further protein in output. But, both the mRNA are found to be translated, and thus proteins in output as a result of a coordinated response by miRNA. \\
This can be compared with the concept of resource competition in biological systems. The total miRNA pool can be considered as a pool of post-transcriptional resources in gene regulation. The miRNA is now simultaneously pairing with both the mRNA pools, depending upon the rate constant and other biological factors. With all equivalent conditions, miRNA can now pair with both the mRNA pool, but practically when a miRNA pairs with an mRNA of any of the pools (say the first pool now), this miRNA is no longer available to pair with other mRNAs of both the pool. Thus, even though the total miRNA is greater than the individual mRNA pool, when a part of the miRNA pool pairs with mRNA, that part is no more available to block the other mRNA. As the sum of mRNA pool is greater than miRNA, total miRNA is segmented, blocking each mRNA pools partially, but letting some active mRNA available for further translation, and both the proteins are found in the output.
\\We study the coordinated response of this resource sharing in the miRNA-based binary gene expression (Model 2), especially to observe the spatio-temporal effect. In the next section, we briefly describe the model formulation part, and then the results. 
\subsection{Model Formulation}
As an extension of Model 2, we consider another mRNA is now simultaneously pairing along with the first pool of mRNA to the miRNA pool. To highlight the effects of resource miRNA sharing, we consider this second mRNA pool (say $m_v$) has no activation via protein as in our previous model and the first mRNA pool (say $m_u$) here has. The second mRNA pool $m_v$ is entirely coupled via the miRNA pool to the first mRNA pool $m_u$. We have established that this resource sharing emerges completely new responses in protein $U$ and $V$ dynamics. 
\subsubsection{Deterministic Model of Coordinated Binary Response}
$U$ and $V$ be the two proteins respectively translated from the two mRNA pool $m_u$ and $m_v$ at rates of $I_u$ and $I_v$. The transcription rates of $m_u$ and $m_v$ are considered as $I_{mu}$ and $I_{mv}$. $m_{iTotal}$ is the total miRNA pool shared between the two mRNA pools ($m_u$ and $m_v$ respectively). $I_{biu}$ is the rate of binding between the accessible free miRNA from the total miRNA pool $m_{iTotal}$ and mRNA $m_u$ to produce a miRNA-mRNA complex, say $m_{cu}$. This complex can further unbind to release the mRNA $m_u$ at a rate $I_{unu}$ which can take part in translation, or degrade with a rate of $\Delta m_{cu}$. The mRNA $m_u$ and protein $U$ can degrade at a rate of $\Delta m_u$ and $\Delta u$ respectively. Similar notations with the suffix $v$ are considered for similar considerations of protein $V$. The first mRNA $m_u$ is considered to be regulated positively by the first protein $U$ with rate constant $\gamma_{mu}$ added with an equilibrium dissociation constant $K_u$, similar to Model 2. However, the second protein $V$ has no such regulation. 
\\Now, to account for the available free miRNA pool, accessible by both the mRNA pool for complex formation from the total miRNA pool $m_{iTotal}$ we follow the conventional way of subtracting the complex miRNA from the total pool. This is the conventional way of determining free resources when shared from a common pool between multiple participants followed by \cite{chakraborty2021emergent,chakraborty2021bemergent,gyorgy2015isocost}. The accessible miRNA pool is thus given by $(m_{iTotal}-m_{cu}-m_{cv})$.
\\The set of differential equations representing above scenario is given in Eqn. \ref{model3}.
\begin{eqnarray}\label{model3}
\frac{dm_u}{dt}= I_{mu}+\frac{\gamma_{mu}\;U}{K_u+U} -I_{biu}\;m_u\;(m_{iTotal}-m_{cu}-m_{cv})+I_{unu}\;m_{cu}-\Delta m_u\;m_u\nonumber\\
 \frac{dm_{cu}}{dt}= I_{biu}\;m_u\;(m_{iTotal}-m_{cu}-m_{cv})\;-I_{unu}\;m_{cu}-\Delta m_{cu}\;m_{cu}\nonumber\\
 \frac{dU}{dt}= I_u\;m_u-\Delta u\;U\nonumber\\
 \frac{dm_v}{dt}= I_{mv} -I_{biv}\;m_v\;(m_{iTotal}-m_{cu}-m_{cv})+I_{unv}\;m_{cv}-\Delta m_v\;m_v\nonumber\\
 \frac{dm_{cv}}{dt}= I_{biv}\;m_v\;(m_{iTotal}-m_{cu}-m_{cv})\;-I_{unv}\;m_{cv}-\Delta m_{cv}\;m_{cv}\nonumber\\
 \frac{dV}{dt}= I_v\;m_v-\Delta v\;V\nonumber\\
\end{eqnarray}
At equilibrium, all the rates of change are equal to zero, and we investigate the model.
\subsubsection{Model Formulation: Reaction-diffusion system:}
We further incorporate diffusion of two proteins $U$ and $V$ for a collection of a two-dimensional sheet of $200\times200$ cells considering tissue layer formation, bio-film generation, etc. in biology. The diffusion coefficient of protein $U$ is considered $D_U^x$ in $x$ direction and $D_U^y$ in $y$ direction and similarly the diffusion coefficient of $V$ is considered as $D_V^x$ for $x$ and $D_V^y$ for $y$ direction. In the presence of diffusion Eqn. \ref{model3} will change to
\begin{eqnarray}\label{model4}
\frac{dm_u}{dt}= I_{mu}+\frac{\gamma_{mu}\;U}{K_u+U} -I_{biu}\;m_u\;(m_{iTotal}-m_{cu}-m_{cv})+I_{unu}\;m_{cu}-\Delta m_u\;m_u\nonumber\\
 \frac{dm_{cu}}{dt}= I_{biu}\;m_u\;(m_{iTotal}-m_{cu}-m_{cv})\;-I_{unu}\;m_{cu}-\Delta m_{cu}\;m_{cu}\nonumber\\
 \frac{\partial U(x,y,t)}{\partial t}= I_u\;m_u-\Delta u\;U +D_U^x\;\frac{\partial^2 U}{\partial x^2}+D_U^y\;\frac{\partial^2U}{\partial y^2}\nonumber\\
 \frac{dm_v}{dt}= I_{mv} -I_{biv}\;m_v\;(m_{iTotal}-m_{cu}-m_{cv})+I_{unv}\;m_{cv}-\Delta m_v\;m_v\nonumber\\
 \frac{dm_{cv}}{dt}= I_{biv}\;m_v\;(m_{iTotal}-m_{cu}-m_{cv})\;-I_{unv}\;m_{cv}-\Delta m_{cv}\;m_{cv}\nonumber\\
 \frac{\partial V(x,y,t)}{\partial t}= I_v\;m_v-\Delta v\;V+D_V^x\;\frac{\partial^2 V}{\partial x^2}+D_V^y\;\frac{\partial^2V}{\partial y^2}\nonumber\\
\end{eqnarray}
We have considered isotropic diffusion here, i.e. for both the proteins $U$ and $V$ have the same diffusion coefficient in both the $x$ and $y$ direction. Thus $D_U^x=D_U^y=D_U$ represents the diffusion coefficient of $U$ and $D_V^x=D_V^y=D_V$ represents the diffusion coefficient of $V$.
\subsection{Results: Bifurcation analysis}
We start with the linear stability analysis and find interesting results in the output. Here, in this Model 3, the first mRNA pool, $m_u$, has the same dynamics of mRNA $m$ of previously discussed Model 2, with the additional effect of sharing resource miRNA pool with another mRNA pool, $m_v$. As we proceed to analyze bifurcation for the current model, we find that the discriminant of the first protein $U$ of Model $3$ has the same form of Eqn. \ref{discriminant} with modified parameters of Eqn. \ref{abbriviatn} as
\begin{equation*}
    \beta^*_{new}= \Delta m_{cu}\;(m_{iTotal}-m_{cv}),\;\;\;\;m_{cu}=\frac{m_u\;(m_{iTotal}-m_{cv})}{m_u+\lambda_a}
\end{equation*}
Thus, similar to Model 2, we observe the bistable nature of protein $U$; but more interestingly here the bistability is affected by the coupling of the second mRNA pool, $m_v$ with miRNA, and also with the availability of the total miRNA pool, $m_{iTotal}$. The second protein $V$ shows a Saddle-node bifurcation, in the presence of miRNA coupling, even though it has simple monostable dynamics in the absence of coupling. We have elaborated on the important findings below. 
\subsubsection{Bistability of Protein $U$ depends upon the binding of miRNA to both the mRNA pools:}
\begin{figure}
    \centering
    \includegraphics[width=\textwidth]{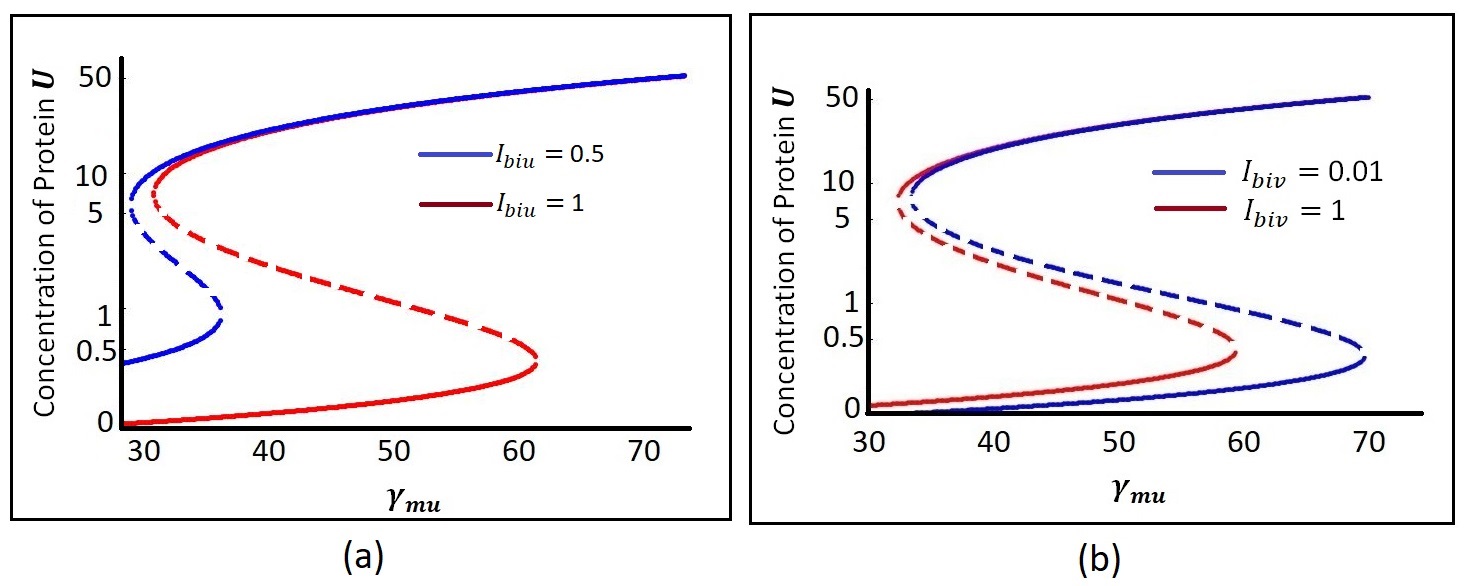}
    \caption{Bistability of protein $U$ depends on the binding rate of miRNA to both the mRNA pools. (a) The concentration of protein $U$ wrt. activation constant $\gamma_{mu}$ plot for two different binding constants of mRNA $m_u$ to miRNA. $I_{biu}=0.5$ for blue curve, $I_{biu}=1$ for red curve. (b) The concentration of protein $U$ wrt. activation constant $\gamma_{mu}$ plot for two different binding constants of mRNA $m_v$ to miRNA. $I_{biv}=0.01$ for blue curve, $I_{biv}=1$ for red curve. The parameter values are $m_{iTotal}=10,\; I_{mu}=I_{mv}=1,\; I_{unu}=I_{unv}=0.1,\; K_u=12,\;I_u=I_v=0.5,\;\Delta m_u=\Delta m_{cu}= \Delta u=\Delta m_v=\Delta m_{cv}= \Delta v=0.7$ for both the model. $I_{biv}=1$ for (a) and $I_{biu}=1$ for (b). }
    \label{Ubistable}
\end{figure}

The bistable behavior of protein $U$, as a result of the miRNA-based post-transcriptional regulation, is found to be affected by the coupling of miRNA to both the mRNA pool, as shown in Fig. \ref{Ubistable}. The binding constant $I_{biu}$ of mRNA $m_u$ with the miRNA pool will regulate the bistability of protein $U$ is a straightforward conclusion of our model, which can be seen in Fig. \ref{Ubistable}(a). Increase in the region of bistability as well as a shift in the point of bifurcation with an increase in $I_{biu}$ ($I_{biu}=0.5$ for blue curve and $I_{biu}=1$ for red curve) is observed. Further, the bistability of protein $U$ is found to be dependent on the binding rate $I_{biv}$ of miRNA to the second mRNA pool $m_v$. Interestingly, $I_{biv}$ is not a parameter regulating the dynamics of protein $U$ directly, this dependency is completely regulated by indirect coupling of the two mRNA pool $m_u$ and $m_v$ via the miRNA resource pool. With all other parameters fixed, a plot of the concentration of protein $U$ wrt. the activation constant $\gamma_{mu}$, for two different values of binding constant ($m_v$ with the miRNA pool) $I_{biv}$, we find a shift in the bistable region along with a change in the region of bistability. The red curve for $I_{biv} = 1$ has a smaller bistable region, starting with bifurcation a little earlier when compared with the blue curve of $I_{biv} = 0.01$.
\subsubsection{Emergent bistability in Protein $V$: Graded vs. Binary response}
\begin{figure}
    \centering
    \includegraphics[width=\textwidth]{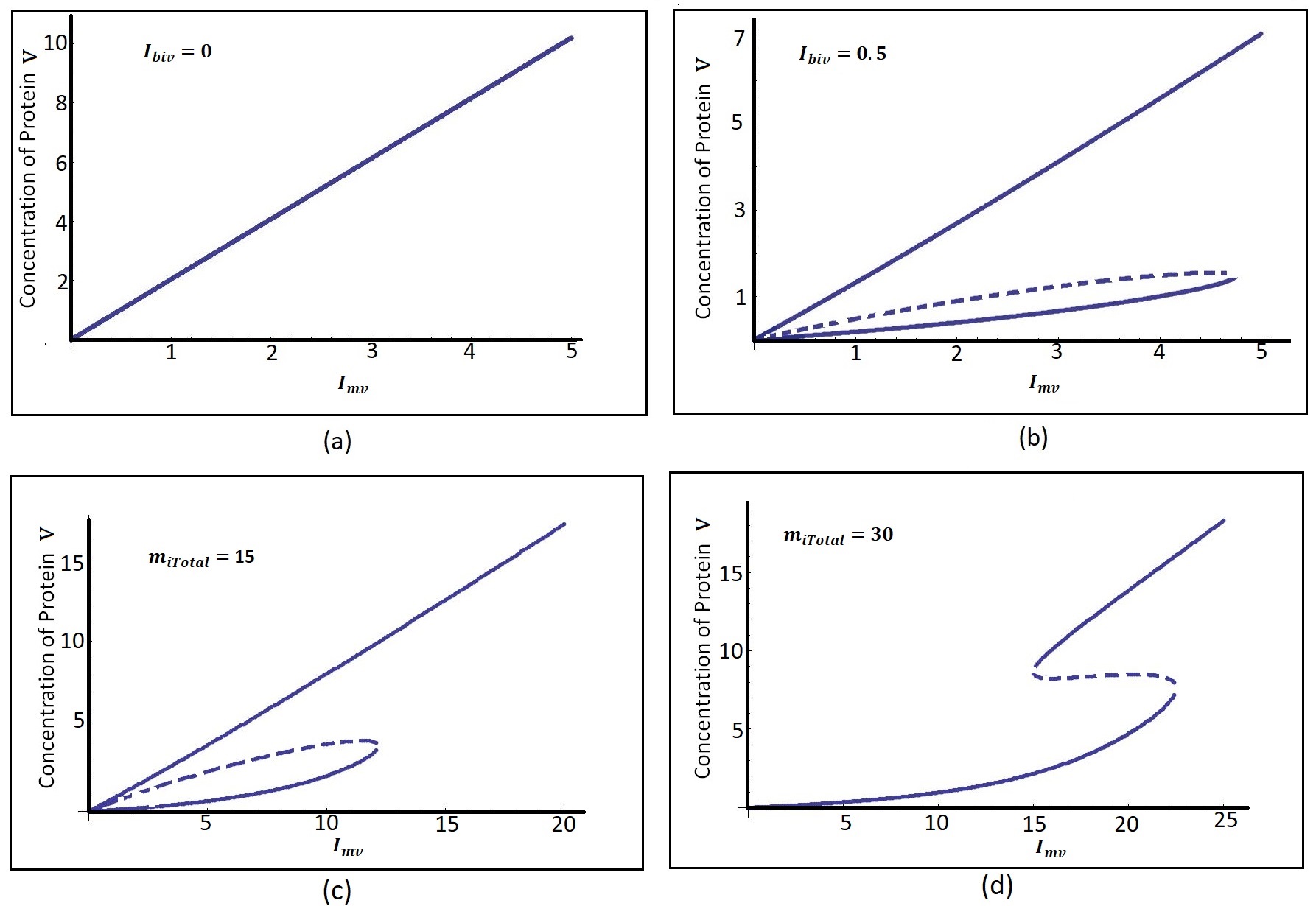}
    \caption{ Emergent bistability in protein $V$, due to sharing same miRNA pool with $U$. (a) In the absence of miRNA binding ($I_{biv}=0$) protein $V$ shows no bistability. (b) Protein $V$ shows bistability in the presence of coupling to miRNA pool with $I_{biv}=0.5$. Parameter values are $m_{iTotal}=20,\;  K_u=3,\;I_v=1,\;\gamma_{mu}=30,\;$for (a) and (b). (c),(d) Bistability of protein $V$ wrt $I_{mv}$ for two different  values of $m_{iTotal}$. $K_u=12,\;I_v=0.5,\;I_{biv}=0.5,\;\gamma_{mu}=50,$ for (c) and (d). $I_{biu}=1,\;I_{mu}=1,\; I_{unu}=I_{unv}=0.1,\;\;I_u=0.5,\;\Delta m_u=\Delta m_{cu}= \Delta u=\Delta m_v=\Delta m_{cv}= \Delta v=0.7$ for all (a)-(d).}
    \label{Vbistable}
\end{figure}

The indirect coupling of protein $V$ with protein $U$, via the resource pool of miRNA significantly modifies the dynamics of protein $V$. Emergent bistability in the dynamics of $V$ is shown in Fig. \ref{Vbistable}. To establish that the bistability in protein $V$ is completely emergent due to sharing a common miRNA pool with $U$, we have studied the dynamics of $V$ in the presence and absence of this resource coupling. In the absence of the coupling (with $I_{biv}=0 $) a monostable dynamics is seen (Fig. \ref{Vbistable}(a)), while in the presence of resource miRNA coupling (with $I_{biv}=0.5$), we find bistability in $V$, shown in Fig. \ref{Vbistable}(b). Graded response of protein $V$, switches to a binary response in the region of bistability due to coupling to $U$ via resource pool. Another important observation here is the change in the shape of bistability curve of protein $V$ with the change in miRNA pool value $m_{iTotal}$ shown in Fig. \ref{Vbistable}(c), (d).  
\subsection{Availability of total miRNA regulates bifurcation of both the proteins: }
\begin{figure}
    \centering
    \includegraphics[width=\textwidth]{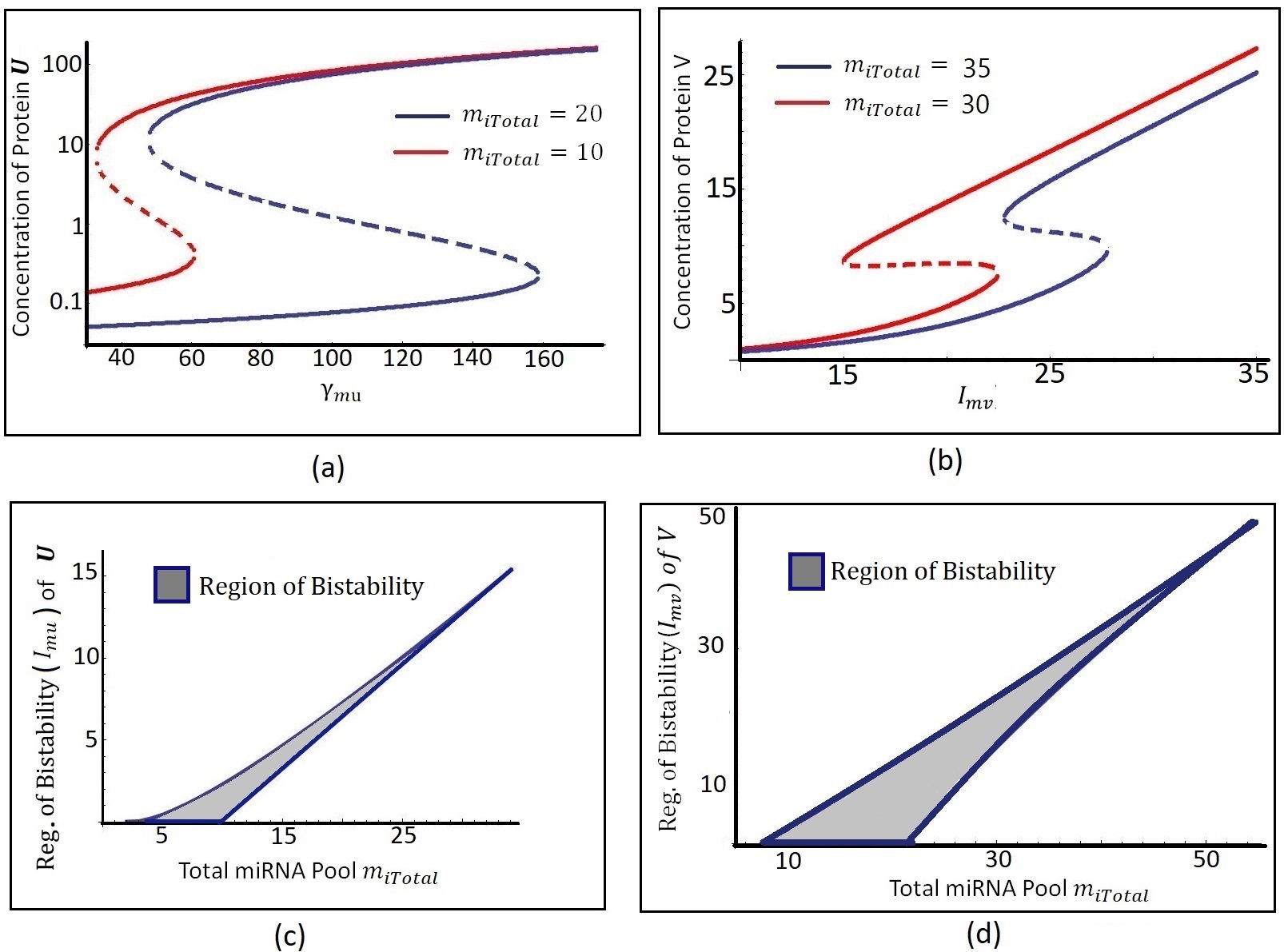}
    \caption{Availability of total miRNA regulates the bifurcation of both proteins. (a) Bistability of protein $U$ depends upon the total availability of miRNA. The concentration of protein $U$ vs. activation constant $\gamma _{mu}$ plot. The red curve is for $m_{iTotal}=10$, the blue curve is for $m_{iTotal}=20$. (b)  Bistability of protein $V$ depends upon the total availability of miRNA. The concentration of protein $V$ vs. transcription rate $I_{mv}$ plot. The red curve is for $m_{iTotal}=30$, the blue curve is for $m_{iTotal}=35$. (c) Phase space plot of protein $U$ for the region of bistability $I_{mu}$ for different $m_{iTotal}$. (d) Phase space plot of protein $V$ for the region of bistability $I_{mv}$ for different $m_{iTotal}$. Parameter values are $K_u=12,\;I_{mu}=1,\;I_{biv}=0.5,\; I_{biu}=1,\; I_{unu}=I_{unv}=0.1,\;\;I_u=I_v=0.5,\;\Delta m_u=\Delta m_{cu}= \Delta u=\Delta m_v=\Delta m_{cv}= \Delta v=0.7$ for all (a)-(d). For (a) $I_{mv}= 1$, For (b),(c),(d) $\gamma_{mu}=50$. }
    \label{total mirna}
\end{figure}

As the coupling between two proteins is completely regulated by sharing the same miRNA pool, the availability of total miRNA $m_{iTotal}$ significantly regulates the bistability of both the proteins $U$ and $V$ as shown in Fig. \ref{total mirna}. With an increase in miRNA pool $m_{iTotal}$, for protein $U$ we find the range of bistability increases considerably, when plotted against the activation constant $\gamma_{mu}$, comparing red and blue curves in Fig. \ref{total mirna}(a). For protein $V$  we plot the concentration of $V$ against the rate of transcription $I_{mv}$. The range of bistable region is found to increase with $m_{iTotal}$ initially then decreases after a certain value of $m_{iTotal}$.  A similar response is found for $U$ also, when plotted wrt. transcription rate of mRNA $m_u$, the parameter $I_{mu}$ (Not shown here). The phase space plot of the protein $U$ in Fig. \ref{total mirna}(c) (phase plot of protein $V$ in Fig. \ref{total mirna} (d)) shows that with the increase in $m_{iTotal}$, we got a closed bistable phase space shown in blue-gray color representing the region of bistability for the rate of transcription $I_{mu}$ of protein $U$ (for the rate of transcription $I_{mv}$ for protein $V$) with a defined blue boundary, which increases initially with the increase in $m_{iTotal}$ up to a certain value and decreases afterward. 
\\The coupling of miRNA to mRNA actually stops the translation, thus for smaller values of $m_{iTotal}$, if the transcription rate of mRNA is low, the miRNA will block most of mRNA and no proteins will be synthesized. The two proteins $U$ and $V$ show similar responses in phase space because of sharing of the same miRNA pool, which reflects coordinated behavior. We can see for a range of lower values of $m_{iTotal}$, bistability starts from a very low value of the transcription rate. When explained in terms of concentration, Fig. \ref{Vbistable}, we can conclude that miRNA coupling includes another low synthesis steady state in the potential along with its high synthesis steady state, which exists even without coupling. When $m_{iTotal}$ is low, for both low values of $I_{mu}$ and $I_{mv}$ and coordinated response of $m_u$ and $m_v$, both the proteins can have a high synthesis state or low synthesis state because of emergent bistability of miRNA-mRNA coupling. Increasing $I_{mu}$ in $U$ (similarly $I_{mv}$ in $V$) for a low value of $m_{iTotal}$ produces many mRNAs, causing very little post-transcriptional regulation and the bistability effect is not prominent by high transcription, thus a translation of proteins. Further, when $m_{iTotal}$ is high, for a low transcription rate of both the proteins $U$ and $V$, ($I_{mu}$ and $I_{mv}$ ), respective mRNAs are repressed so resulting in very low protein synthesis. Bistability is seen for intermediate values of transcription rates. For a very high value of $m_{iTotal}$ most of the mRNAs are repressed and bistability is seen for a small region of high transcription values. Thus a closed bistable phase space in the output is seen. 
\\Coordinated actions of miRNA is well known to regulate different significant cellular decision processes, which include regulation of skeletal muscle development and adaptation \cite{bianchi2017coordinated}, a transition from epithelial to mesenchymal (EM) or mesenchymal to epithelial (ME) state of cells in cancer systems \cite{cursons2018combinatorial} and many more. Here, in our model we find the coordinated response, as a result of miRNA sharing from a fixed pool by two separate mRNA pools ($m_u$ and $m_v$), emerges bistability in the otherwise monostable dynamics of protein $V$. Instead of the graded linear response the switch-like response of protein $V$ with a memory, driven by the saddle-node bifurcation in the system, significantly changes the protein concentration. Biologically this is very impactful as the genetic networks are mostly complex and dependent upon each other's expression. Thus the emergent bistability of protein $V$, and thus the binary concentration of $V$, can affect the local dynamics and also the global dynamics via the connected genetic network as well.   
\subsection{Result: Reaction-diffusion and Pattern formation}
\subsubsection{Transient pattern formation in protein $V$ due to miRNA coupling:}
\begin{figure}
    \centering
    \includegraphics[width=\textwidth]{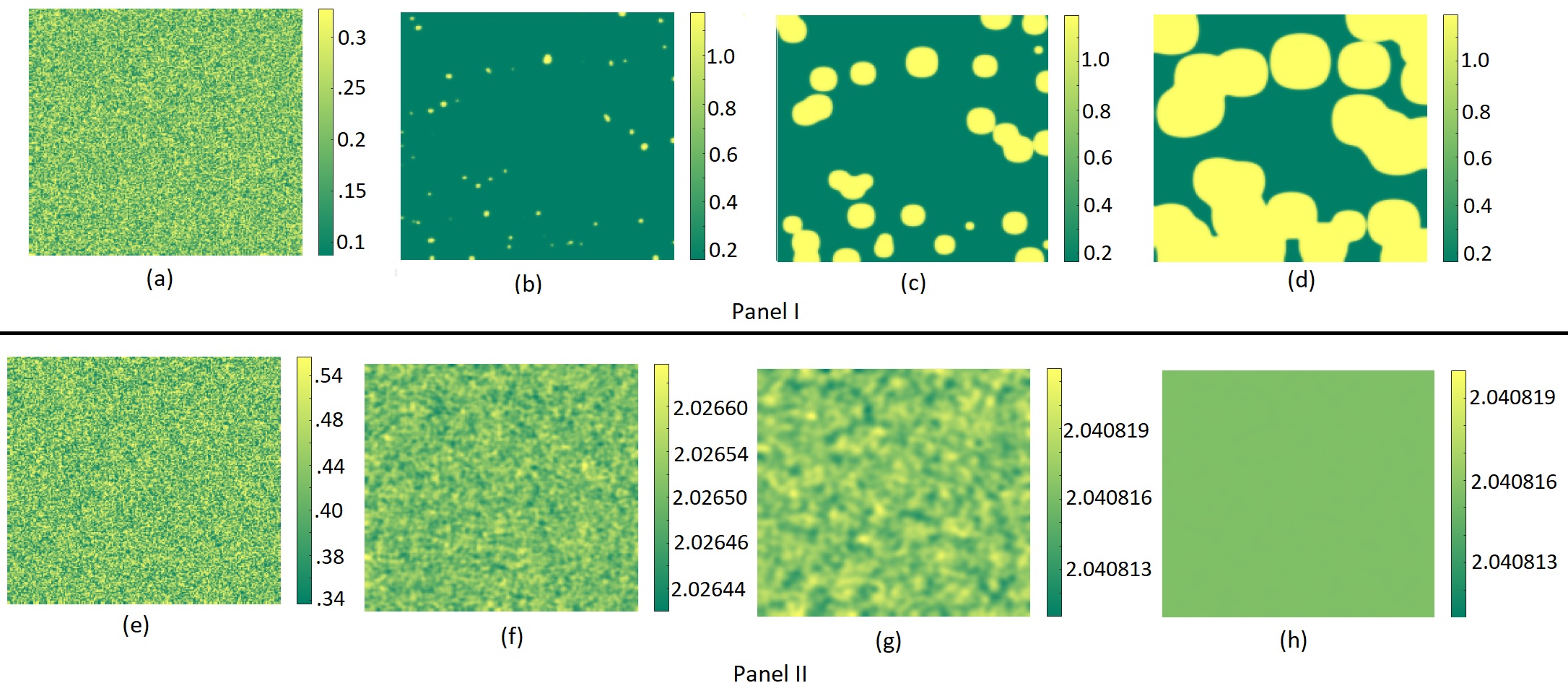}
    \caption{Resource sharing emerges transient pattern formation in protein $V$. Panel I: In the presence of miRNA binding with protein $V$, a transient pattern is seen in a two-dimensional sheet of cells. Snapshots are taken after time (a) 1, (b) 45, (c) 250, (d) 470. Panel II: In the absence of miRNA binding, no pattern is seen in a two-dimensional sheet of protein $V$. Snapshots are taken after time (e) 1 (f) 10 (g) 15 (h) 20. Diffusion coefficients are taken as  $D_U=0.1,\;D_V=0.1$ Parameter values are $k_1=10,\;k_2=0.5,K_u=3,\;m_{iTotal}=22,\;I_{mu}=I_{mv}=1,\;I_{biu}=1,\; I_{unu}=I_{unv}=0.1,\;I_u=0.5,\;I_v=1,\;\gamma_{mu}=30,\;\Delta m_u=\Delta m_{cu}= \Delta u=\Delta m_v=\Delta m_{cv}= \Delta v=0.7$ for all the figures of (a) to (h). For Panel I (a)-(d) $I_{biv}=0.5$, for panel II (e)-(h) $I_{biv}=0$.}
    \label{pattern-coupling}
\end{figure}

The emergent bistability in the dynamics of protein $V$, due to the coupling with protein $U$ via the miRNA pool, further causes transient pattern formation in presence of diffusion, when studied in a two-dimensional sheet of cells. In presence of coupling (with $I_{biv}=0.5$), when started from an initial randomized condition:

\begin{equation*}
U_{initial}(x_i,y_i,0)=k_1\;\xi(0,1)\;\;\;\;V_{initial}(x_i,y_i,0)=k_2\;\xi(0,1)    
\end{equation*}
each cell chooses any of the steady states (either low or high) and the bistable behavior causes a binary response in output (also shown in Fig. \ref{pattern-coupling}(a), Panel I). The initial condition is randomized as chosen above by considering a factor $\xi$ to pick up any number between $0$ to $1$, multiplied with a scaling factor $k_1$. Due to diffusion, we can further see a transition of low to high expression state causing a  transient pattern in output and finally evolving the entire sheet to the high expression state (Fig. \ref{pattern-coupling}(a)-(d), Panel I). In time evolution appearance of island-like patches gives spatio-temporal richness to the dynamics.
\\ In the absence of miRNA coupling (with $I_{biv}=0$), no such binary response and further pattern formation are seen (Fig. \ref{pattern-coupling}(e)-(f), Panel II). Starting from the same randomized condition, the system quickly evolves to its single steady state, without any visible pattern in output.
This shows that this coupling of mRNA $m_v$ with the miRNA pool $m_{iTotal}$ changes the dynamics of protein $V$ in steady state and reaction-diffusion approach as well. This emergent transient pattern might be a reason for phenotypic heterogeneity when shedding impact to its nearby genetic network.   
\subsubsection{Pattern Formation from Initial Homogeneity:}
\begin{figure}
    \centering
    \includegraphics[width=0.8\textwidth]{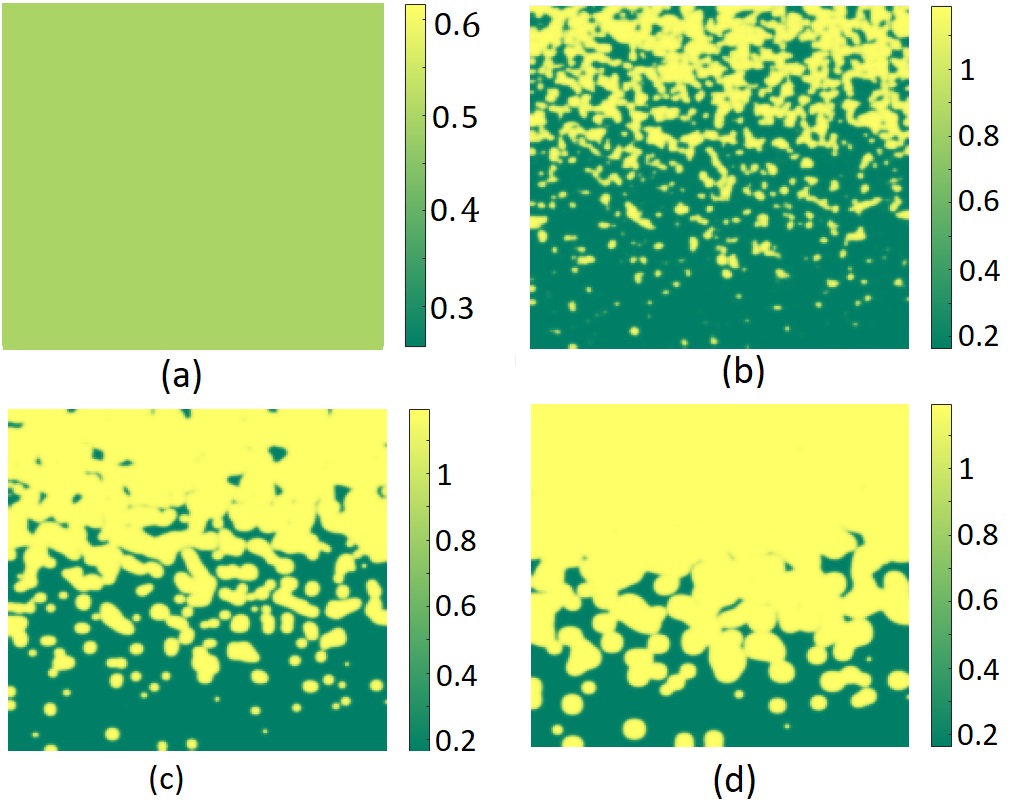}
    \caption{Pattern formation in protein $V$ from homogeneous initial state. Snapshots are taken for protein $V$ after time (a) 0 (b) 55 (c) 75 (d) 135. Diffusion coefficients are taken as $D_U=0.1,\;D_V=0.1$. Parameter values are $k_1=10,\;k_2=0.5,\;k_3=0.0005,K_u=3,\;m_{iTotal}=22,\;I_{mu}=I_{mv}=1,\;I_{biu}=1,\;I_{biv}=0.5,\; I_{unu}=I_{unv}=0.1,\;I_u=0.5,\;I_v=1,\;\gamma_{mu}=30,\;\Delta m_u=\Delta m_{cu}= \Delta u=\Delta m_v=\Delta m_{cv}= \Delta v=0.7$.}
    \label{exp in v}
\end{figure}

The coupling of protein $U$ to $V$ via the miRNA pool causes emergent bistability in dynamics of $V$, and pattern formation considering diffusion in a two-dimensional sheet of cells. This coupling is so effective that an initial pre-patterning in $U$ can surpass the initial homogeneity conditions for $V$, eventually dictating an instability and pattern formation of the $V$. For example, suppose, $U$ is initialized with an exponential form as shown below:
\begin{equation*}
U_{initial}(x_i,y_i,0)=k_1\;\xi(0,1)\;exp(\pi\;x_i\;k_3)\;\;\;\;V_{initial}(x_i,y_i,0)=k_2    \end{equation*}
while, the initialization of $V$ is fixed and homogeneous. Though $V$ initialization is fixed in a constant value, the initial pre-pattern in protein $U$ accordingly couples with miRNA; thus, a shadow of $U$'s pattern is left in $m_{iTotal}$ pool. Further, this spatial distribution in $m_{iTotal}$, via the coupling with $m_v$ patterns $V$ spatially. Thus, instability arises creating islands and patches for protein $V$. As a result, an exponential evolution of protein $V$ with time in the two-dimensional sheet of cells can be seen in Fig. \ref{exp in v}. This shows the possibility of patterning in a static cell lattice, which remarkably drifts away from a stable homogeneous state due to miRNA-based regulatory coupling.  
\section{Discussion}
A combination of reaction-driven stability and diffusion-driven instability (that causes an otherwise stable spatial state to become unstable) may result in pattern formation for biological systems. Biological systems naturally fall under the category of reaction-diffusion systems where reaction serves the purpose of intercellular dynamical actions, and diffusion establishes intracellular signaling and environmental communication. It is considered that reaction-diffusion causes pattern formation in biological systems like animals, birds, fish, and many more in a wide aspect of shapes (dotted, periodic, strips etc.) and colors. The study of spatio-temporal pattern formation in biological systems has unravelled the dynamics of development and phenotypic heterogeneity as well. Scientists are recently focusing on exploring synthetically tunable pattern formations theoretically and experimentally, as models can be the best ways to take steps towards studying natural systems, which are tremendously complicated because of numerous confounding factors. Further insight regarding pattern formation as an emergent phenomena will reveal the self-organization of cells in three dimensions, bio-film development,  quorum sensing in pathogens, and many more. 
\\Here, we have studied three motifs of miRNA-regulated mRNA synthesis, respective dynamics, and spatio-temporal pattern formations. Though the role of miRNAs in spatially heterogeneous time evolution is well-known, it still remains a less explored domain from mathematical and computational perspectives.  To understand phenomena like development, it is actually necessary to consider the transient changes in the dynamics rather than focus on the final state. Sluggish transient patterns are completely capable of initiating important decision sequences. However, the transient pattern formations are generally ignored in the case of pattern formation studies. We have studied three scenarios (all parameters and their physical significance have been consolidated in Appendix, Table 1,2,3.) where miRNA-regulated gene expression creates spatial patterning, which are sustainable enough to dictate biological decisions. We start by exploring the silencing of gene expression by post-transcriptional regulation, or more specifically by the absence/presence of miRNAs, providing a threshold in gene expression. This provides a clear idea of environmental fluctuations, and stress management in biological systems and we have demonstrated the spatial implications of threshold response with the results for Model 1.
\\In Model 2, we have explored a bistable motif, arising due to auto-activation of the gene of interest, under miRNA regulation. The steady-state analysis, along with a brief reaction-diffusion analysis, explored the possible dynamics of the system in natural or synthetic environments. Some transient spatio-temporal patterns are shown, including a scenario of convergence to a particular steady state and scenarios of long persistent patterns much away from converging to a particular state. There is a chance of these patterns getting stable in the cellular environment in the presence of different biological factors, which is not included explicitly in our computational model. The transient patterns are also very important in the light of development, influencing the cellular decision-making process for its own fate as well as for the connected genetic network.  
\\In Model 3, we have explored a competitive scenario of miRNA between two mRNA pools as an extension of our Model 2, adding a new mRNA pool, regulated by miRNA from the same pool. The coordinated response of two mRNA allows the expression of both pools, giving rise to some novel behavior in the dynamics. The most striking observation shows coupling-induced emergent bistability in the dynamics of the second protein, which is temporally as well as spatially exhibited. The coordinated response in post-transcriptional regulation causes transient patchy pattern formation in a protein, causing regional differences in the concentrations, starting from a homogeneously distributed initial condition.
\\We have also considered the fact that different environmental signaling cues may pre-pattern the system, in terms of initial conditions. To incorporate different distribution profiles, we have considered different initial conditions along with random initialization, as in de-novo pattern formation, a prior step of pattern formation is the absence of spatial information. We have considered symmetric diffusion considering the same diffusion coefficient in both directions now. The work can be extended further for asymmetric diffusion coefficients and diffusion gradients. As in biological systems protein molecules inherently differ in diffusion coefficient in a different direction and also might be a function of space. 
In some recent studies, the variation in transient time with the change in system size \cite{crutchfield1988attractors}, the regulation of temperature and evaporation in pattern dynamics, more accurately in the complexity of the pattern \cite{hamann2012self} opens a new insight of further theoretical and experimental study of our systems.  We also believe scenarios related to diseases where dysregulation of miRNA has been observed can be better understood and explored through this study in the future. 
\section*{Conflict of Interest}
The authors declare that they do not have any known conflicts of interest.
\section*{Acknowledgement}
\noindent PC and SG acknowledge the support  by DST-INSPIRE, India, vide sanction Letter No. 
DST/INSPIRE/04/2017/002765 dated- 13.03.2019.
\section*{Data Availability}
\noindent The manuscript has no associated data.
\bibliographystyle{abbrv}
\bibliography{bibliography.bib}
\newpage
\section*{Appendix}
\subsection*{List of parameters:}
For better clarity, we have given a list of parameters used in our entire manuscript hereby in Tables 1, 2, 3. 
\begin{table}[!b]
\begin{center}
\begin{tabular}{ |p{4cm}|p{5cm}|  }

 \hline
 Parameter Name & Physical Meaning\\
 \hline
 S  & miRNA   \\
 \hline
 m & mRNA  \\
 \hline
 P & Protein\\
 \hline
 $\alpha_s$ & miRNA transcription rate\\
 \hline
 $\alpha_m$ & mRNA transcription rate\\
 \hline
 $\alpha_p$ & The synthesis rate of protein\\
 \hline
 $\beta_s$ & Natural degradation rate of miRNA\\
 \hline
 $\beta_m$ & Natural degradation rate of mRNA\\
 \hline
 $\delta$ & Natural degradation rate of protein\\
 \hline
 k & Rate of complex formation between miRNA and mRNA\\
 \hline
 $D_P^x$ & Rate of protein diffusion along x axis\\
 \hline
 $D_P^y$ & Rate of protein diffusion along y axis\\

 \hline
\end{tabular}
\caption{List of parameters for Model 1}
\end{center}
\end{table}

\begin{table}
\begin{center}
\begin{tabular}{ |p{4cm}|p{6cm}|  }

 \hline
 Parameter Name & Physical Meaning\\
 \hline
 m & mRNA  \\
 \hline
 mi, $mi_{Total}$, $m_c$ & Free miRNA, Total pool of available miRNA, miRNA-mRNA complex\\
 \hline
 $U$ & Protein\\
 \hline
 $\gamma_m$ & Auto activation rate constant for mRNA\\
 \hline
 $I_m$ & Transcription rate of mRNA\\
 \hline
 $I_{bi}$ & Binding coefficient of miRNA-mRNA to produce complex\\
 \hline
 $I_{un}$ & Unbinding coefficient of the complex\\
 \hline
 $I_u$ & Rate of protein U production\\
 \hline
 k & Equilibrium dissociation constant\\
 \hline
 $\Delta m$, $\Delta m_c$, $\Delta u$ & The degradation rate of mRNA, miRNA-mRNA complex, protein.\\
 \hline
 
\end{tabular}
\caption{List of parameters for Model 2}
\end{center}
\end{table}

\begin{table}
\begin{center}
\begin{tabular}{ |p{2cm}|p{3cm}|p{2cm}|p{3cm}|  }

 \hline
 Parameter Name & Physical Meaning &Parameter Name & Physical Meaning\\
 \hline
 U & First type of protein (say) with auto-activation in mRNA production & V & Second type of protein (say) with no auto-activation in mRNA production  \\
 \hline
 $m_u$ & mRNA of protein U & $m_v$ & mRNA of protein V\\
 \hline
 $m_{cu}$ & miRNA-mRNA complex of protein U & $m_{cv}$ & miRNA-mRNA complex of protein V\\
 \hline
 $\gamma_{mu}$ & Auto activation rate constant for mRNA $m_u$ & $K_u$ & Equilibrium dissociation constant for protein U\\
 \hline
$I_{biu}$ & Binding constant between miRNA and mRNA  $m_{u}$ & $I_{biv}$ & Binding constant between miRNA and mRNA $m_{v}$ \\
\hline
$I_{unu}$ & Unbinding constant between miRNA and mRNA complex $m_{cu}$ & $I_{unv}$ & unbinding constant between miRNA and mRNA complex $m_{cv}$ \\
\hline
$\Delta m_{u}$, $\Delta m_{cu}$, $\Delta u$ & Degradation rate of mRNA $m_u$, miRNA-mRNA complex $m_{cu}$, protein $U$ & $\Delta m_{v}$, $\Delta m_{cv}$, $\Delta v$ & Degradation rate of mRNA $m_v$, miRNA-mRNA complex $m_{cv}$, protein $V$ \\
\hline
$I_u$ & Rate of U production & $I_v$ & Rate of V production\\
\hline
\end{tabular}
\caption{List of parameters for Model 3}
\end{center}
\end{table}

\newpage

\subsection*{Model 2: Binary gene expression, temporal dynamics}
\begin{figure}
    \centering
    \includegraphics[width=\textwidth]{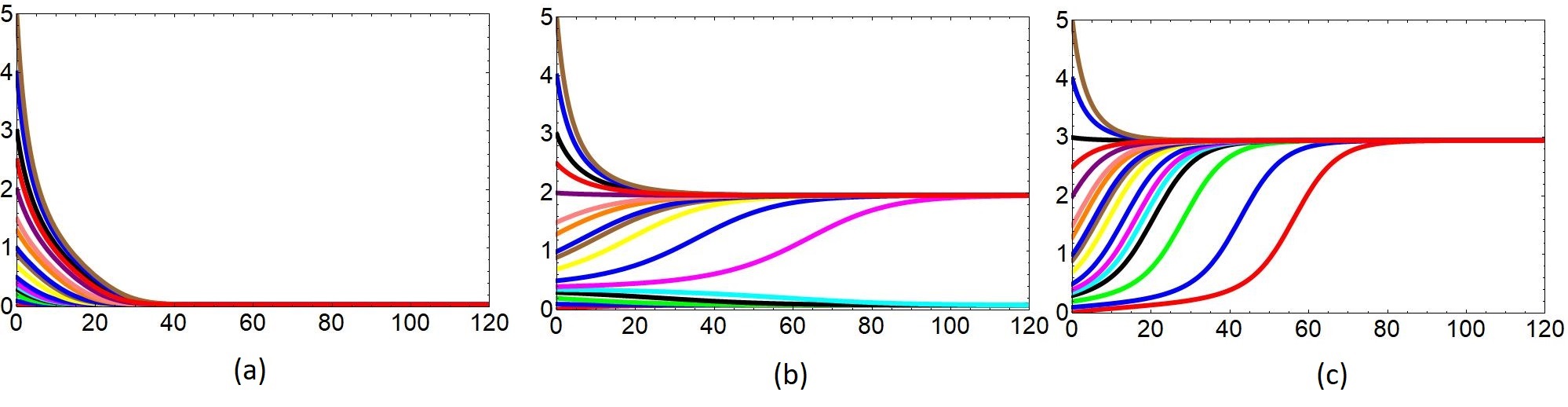}
    \caption{Time evolution curves of protein $U$ for different initial conditions. $x$ axis represents time and $y$ axis represents the concentration of protein $U$. (a) The system is monostable with low synthesis state as a steady state. So for all initiation, protein $U$ goes to its low synthesis stable state. (b) The system is bistable with two steady states. So with different initial conditions, the system chooses any of its nearest stable states and finally, we are left with two stable fixed states. (c) The system is monostable with its high synthesis stable state. All states from different initial conditions move to a single high synthesis stable state. Parameter $\alpha$ has value $12$ for (a), $13.3$ for (b), $14$ for (c). Rest of the parameter values for all (a), (b), (c) are $\lambda=10,\; k=10,\;\delta=0.01,\;\phi=0.3$. }
    \label{flowline}
\end{figure}

To understand the temporal dynamics of the protein synthesized, we plot the time evolution curves of the protein starting from different initial conditions for three set of parameter values. For the parameter values of Fig. \ref{flowline}(a) low synthesis state is the system's stable steady state, starting from all initialization, the system converges to it. Similarly, \ref{flowline}(c) can be explained for its high synthesis stable state. However, the system has two steady states for the parameter value of \ref{flowline}(b), and a bistable dynamics is shown in the output. Starting from different initial concentrations, the protein chooses any of its either low or high synthesis states, which one is more favorable and two drastic different concentrations coexist in output.  



\end{document}